\documentclass[pra,twocolumn,showpacs,superscriptaddress,floatfix, nofootinbib]{revtex4-2}
\usepackage[caption=false]{subfig}
\usepackage{amsmath,graphicx}
\usepackage{amsmath,amssymb,mathrsfs,esint}
\usepackage{multirow}
\usepackage{algorithm}
\usepackage[noend]{algpseudocode}
\usepackage{comment}
\usepackage{tabularx}
\usepackage{bm}
\usepackage[normalem]{ulem}
\usepackage[bookmarks=true,
   colorlinks=true,
   linkcolor=blue,
   urlcolor=blue,
   citecolor=blue,
   bookmarks=true,
   hyperindex=true
]{hyperref}
\usepackage{svg}
\usepackage{xifthen}
\usepackage[english, russian]{babel}
\usepackage{amsmath}
\usepackage{physics}

\newcommand{\iswap}{\ensuremath{\mathtt{iSWAP}}}

\begin{document}

\title{Progress in the development of quantum algorithms and software}

\author{Anastasiia S. Nikolaeva}
\affiliation{Russian Quantum Center, Skolkovo, Moscow 121205, Russia}
\affiliation{National University of Science and Technology ``MISIS”,  Moscow 119049, Russia}
\affiliation{P.N. Lebedev Physical Institute of the Russian Academy of Sciences, Moscow 119991, Russia}

\author{Daria O. Konina}
\affiliation{Russian Quantum Center, Skolkovo, Moscow 121205, Russia}
\affiliation{P.N. Lebedev Physical Institute of the Russian Academy of Sciences, Moscow 119991, Russia}

\author{Anatolii V. Antipov}
\affiliation{Russian Quantum Center, Skolkovo, Moscow 121205, Russia}
\affiliation{National University of Science and Technology ``MISIS”,  Moscow 119049, Russia}

\author{Maksim A. Gavreev}
\affiliation{Russian Quantum Center, Skolkovo, Moscow 121205, Russia}
\affiliation{National University of Science and Technology ``MISIS”,  Moscow 119049, Russia}

\author{Konstantin M.~Makushin}
\affiliation{Russian Quantum Center, Skolkovo, Moscow 121205, Russia}
%\affiliation{Казанский Федеральный Университет, Казань, 420008, Россия}

\author{Boris I. Bantysh}
\affiliation{Russian Quantum Center, Skolkovo, Moscow 121205, Russia}

\author{Andrey Yu. Chernyavskiy}
\affiliation{Russian Quantum Center, Skolkovo, Moscow 121205, Russia}

\author{Grigory V. Astretsov}
\affiliation{Russian Quantum Center, Skolkovo, Moscow 121205, Russia}

\author{Evgeniy A. Polyakov}
\affiliation{Russian Quantum Center, Skolkovo, Moscow 121205, Russia}

\author{Aidar I. Saifoulline}
\affiliation{Russian Quantum Center, Skolkovo, Moscow 121205, Russia}

\author{Evgeniy O. Kiktenko}
\affiliation{Russian Quantum Center, Skolkovo, Moscow 121205, Russia}
\affiliation{National University of Science and Technology ``MISIS”,  Moscow 119049, Russia}

\author{Alexey N. Rubtsov}
\affiliation{Russian Quantum Center, Skolkovo, Moscow 121205, Russia}

\author{Aleksey K. Fedorov}
\affiliation{Russian Quantum Center, Skolkovo, Moscow 121205, Russia}
\affiliation{National University of Science and Technology ``MISIS”,  Moscow 119049, Russia}

\begin{abstract}
	A quantum processor, like any computing device, requires the development of both hardware and the necessary set of software solutions, starting with quantum algorithms and ending with means of accessing quantum devices. As part of the roadmap for the development of the high-tech field of quantum computing in the period from 2020 to 2024, a set of software solutions for quantum computing devices was developed. This software package includes a set of quantum algorithms for solving prototypes of applied tasks, monitoring and benchmarking tools for quantum processors, error suppression and correction methods, tools for compiling and optimizing quantum circuits, as well as interfaces for remote cloud access. This review presents the key results achieved, among which it is necessary to mention the execution of quantum algorithms using a cloud-based quantum computing platform.
\end{abstract}

\maketitle

\section{Введение}

Одним из ключевых факторов роста интереса к квантовым вычислениям стали работы, продемонстрировавшие возможность получения значительного ускорения при решении ряда востребованных задач, таких как разложение на простые множители (имеет прямые приложения в области криптоанализа) \cite{Shor1994}, моделирования сложных квантовых систем \cite{Lloyd1996}, поиска по неупорядоченным базам данных \cite{Grover1997} и др. Благодаря этим исследованиям стало очевидно, что разработка масштабируемого квантового процессора откроет возможности для различных направлений промышленности \cite{Fedorov2022}. За последние десятилетия значительный прогресс был достигнут в части разработки прототипов квантовых процессоров на различных физических принципах: ионах в ловушках \cite{Monroe2017,Blatt2012,Blatt2018, zalivako2025ufn}, нейтральных атомах \cite{Lukin2021,Browaeys2021,Browaeys2020-2,Saffman2022}, сверхпроводниковых цепях \cite{Martinis2019,Pan2021}, оптических системах \cite{Pan2020,Lavoie2022}, спинах в полупроводниках \cite{Loss1998,Vandersypen2022,Morello2022,Tarucha2022} и др. Такие устройства уже могут быть использованы для решения прототипов прикладных задач и демонстрации квантового вычислительного преимущества \cite{arute2019}. Однако наличие в них шумов и ограниченное количество кубитов зачастую затрудняет возможность запуска квантовых алгоритмов для решения прикладных задач \cite{Fedorov2022}. Кроме того, для обеспечения доступа к квантовым процессорам необходимо решение комплекса задач, связанного с созданием соответствующего программного обеспечения.  

В рамках Дорожной карты по квантовым вычислениям, реализуемой под руководством Госкорпорации «Росатом» в период с 2020 по 2024 год \cite{Fedorov2019}, был выделен ряд направлений. Во-первых, направление, связанное с разработкой квантовых алгоритмов. Речь идет как о том, чтобы создать библиотеку известных квантовых алгоритмов, которые можно использовать, в том числе, для демонстрации качества работы квантовых процессоров, так и разработать алгоритмы для решения прототипов прикладных задач, например, из области квантовой химии~\cite{Gircha2023,sapova2022variational, Fedorov2022} и оптимизации~\cite{Shaydulin2024}. Начиная с 2023 года начались первые прикладные проекты по разработке квантовых алгоритмов в интересах атомной отрасли~\cite{Usmanov2024, Bozhedarov2024}. Во-вторых, для эффективного запуска квантовых алгоритмов необходимо было разработать набор методов для их эффективной компиляции -- возможности реализовать их наиболее оптимальным образом для каждого конкретного физического процессора. В свою очередь, это выражается в том, что абстрактный квантовый алгоритм, например, в форме квантовой цепочки, преобразуется в формат, содержащий исключительно нативные операции для каждого конкретного процессора, а количество необходимых операций -- минимально. Особый интерес в данном контексте представляют многоуровневые квантовые системы – кудиты \cite{Kiktenko2023rmp}, которые позволяют более эффективно реализовывать квантовые алгоритмы \cite{Nikolaeva2021epj}. 
В-третьих, для оценки корректности работы квантовых процессоров разработан модуль для мониторинга и бенчмаркинга. Среди основных инструментов – квантовая томография состояний и процессов \cite{Kurmapu2023, Lvovsky2009, norkin2024reliable,Kiktenko2021polytopes}. Также была разработана система непрерывного мониторинга квантовых процессоров~\cite{Zolotarev2023}. В-четвертых, исследовались методы коррекции и подавления ошибок. Основной фокус в данных исследованиях связан со снижением ресурсов, требуемых для запуска кодов коррекции ошибок \cite{antipov2023realizing, Antipov2022}. Пятым направлением стало создание эмуляторов квантовых процессоров -- классического программного обеспечения, которое может моделировать квантовое вычислительное устройство до определенного масштаба (до 30 кубитов). Разработаны эмуляторы для универсальных квантовых вычислительных устройств. Наконец, шестое направление связано с созданием набора инструментов удаленного доступа к квантовым вычислительным устройствам и единой программной среды для объединения всего квантового программного обеспечения -- облачной платформы квантовых вычислений. 

В настоящем обзоре подробно описываются вышеупомянутые направления работы и приводятся ключевые достигнутые результаты, среди которых необходимо отдельно отметить запуск квантовых алгоритмов с помощью облачной платформы квантовых вычислений. Также обсуждаются будущие направления для развития квантовых вычислений в части программного обеспечения в период до 2030 года.

\section{Квантовые алгоритмы} \label{sec:quantum_algorithms}

Квантовый алгоритм представляет собой конечную последовательность шагов для получения решения задачи, предназначенную для выполнения с помощью квантовых вычислительных устройств. В наиболее распространенной цифровой (также называемой вентильной или гейтовой) модели квантовых вычислений квантовый алгоритм может быть формализован в виде квантовой цепочки -- набора инструкций в виде конкретных вентилей (гейтов), действующих на набор кубитов (минимальных бинарных информационных единиц в квантовой теории информации). Стоит отметить, что квантовая цепочка содержит в себе вентили (однокубитные, двухкубитные или многокубитные), которые могут не соответствовать конкретным операциям, выполняемым на каком-либо квантовом процессоре, поэтому для запуска её необходимо преобразовать в необходимую форму – компилировать и транспилировать (см. Раздел~\ref{sec:Compilation_and_optimization}). 

В ходе работ в рамках Дорожной карты по квантовым вычислениям были разработаны разнообразные квантовые алгоритмы. 
Во-первых, это блок алгоритмов, выступающих в качестве элементов собственной библиотеки реализованных и протестированных квантовых алгоритмов, которые могут быть использованы, например, для проверки корректности работы того или иного квантового вычислительного устройства, а также для кроссплатформенного сравнения. К таким алгоритмам можно отнести, например, алгоритмы Гровера \cite{Grover1997}, Бернштейна-Вазирани \cite{bv1993} и алгоритм симуляции фазовых переходов с нарушением симметрии «четность-время» \cite{Kazmina2024demonstration}, которые были запущены на ионном и сверхпроводниковом квантовых компьютерах~\cite{zalivako2024qb16,Kazmina2024demonstration}.

Во-вторых, часть квантовых алгоритмов разрабатывалась для решения прототипов прикладных задач, например, из области оптимизации, машинного обучения и квантовой химии. 

Ряд алгоритмов для комбинаторной оптимизации в данный момент исследуется на предмет применимости для решения задач промышленности. В ходе реализации Дорожной карты был разработан алгоритм в интересах нефтегазовой отрасли, а также исследуются применения для атомной промышленности \cite{Usmanov2024, Bozhedarov2024}. Так, например, был разработан {алгоритм дискретной оптимизации для задач планирования}, который решает задачу оптимизации схем перезагрузки активной зоны в ядерных реакторах. Алгоритм позволяет найти оптимальную схему загрузки реактора в течение заданного количества итераций с учетом ограничений. В алгоритме также реализован процесс постобработки найденного решения в случае, если оно не удовлетворяет ограничениям задачи. Процесс постобработки включает в себя улучшение симметрии схемы загрузки и проведение замены элементов, нарушающих ограничения на размещение. Данный подход иллюстрирует способ повышения эффективности рассмотренного класса алгоритмов. 
Ниже мы подробно рассматриваем наиболее интересные классы разработанных алгоритмов; полный перечень алгоритмов, реализованных в рамках Дорожной карты «Квантовые вычисления», приведен в Приложении А. %\ref{add_alg}.

На следующем цикле развития квантовых вычислений особое внимание предполагается уделить решению комбинаторных оптимизационных задач, для которых были за последние годы показаны перспективы получения ускорения даже в условиях шумов \cite{Shaydulin2024}.

\subsection{Алгоритмы семейства квантовой приближенной оптимизации}

Одним из наиболее исследуемых квантовых алгоритмов, имеющих обозримую перспективу практического использования, является семейство \textbf{\textit{алгоритмов квантовой приближенной оптимизации}} (quantum approximate optimization algorithm, QAOA)~\cite{Farhi2014}. Общая схема алгоритма определяется попеременным применением смешивающего гамильтониана $H_{\rm mix}$ и гамильтониана задачи $H_{\rm targ}$:
\begin{equation}\label{eq:qaoa_circ}
    \ket{\beta,\gamma}=\qty[\prod_{j=1}^{p}{e^{-i \beta_j H_{\rm mix}}e^{-i \gamma_j H_{\rm targ}}}] \ket{+}^{\otimes n},
\end{equation}
где $n$ -- число используемых кубитов, $\{\beta_i, \gamma_i\}_{i=1}^p$ для некоторого $p\geq 1$ -- параметры, соответствующе динамике троттеризованной адиабатической эволюции \cite{roland2002quantum} и $\ket{+}\equiv 2^{-1/2}(\ket{0}+\ket{1})$ (здесь и далее $\ket{0}$ и $\ket{1}$ обозначают вычислительный базис одиночного кубита).
В большинстве работ (например, \cite{guerreschi2019qaoa, zhou2020quantum, fernandez2022study}) алгоритм QAOA реализуется следующим образом. В качестве смешивающего гамильтониана берется $H_{\rm mix}=\sum_{k}{\sigma_x^{(k)}}$ (здесь и далее $\sigma_\alpha^{(k)}, \alpha=x,y,z$ обозначает стандартный оператор Паули $\sigma_\alpha$, действующий на $k$-ый кубит), а гамильтониан задачи 
\begin{equation}
    H_{\rm targ}=\sum_{i<j}s_{ij}\sigma_z^{(i)}\sigma_z^{(j)}+\sum_{i}s_{ii}\sigma_z^{(i)}
\end{equation}
напрямую кодирует так называемую задачу квадратичной неограниченной бинарной оптимизации (quadratic unconstrained binary optimization, QUBO): минимизацию функции
\begin{equation}
    f(\mathbf{z}) = \sum_{i<j}s_{ij}z_iz_j + \sum_i{s_{ii}z_i}    
\end{equation}
от набора бинарных переменных $z_i\in\{+1,-1\}$ для некоторой квадратичной формы с коэффициентами $(s_{ij})$. Квантовый процессор используется для расчёта средних значений $E(\gamma,\beta) = \expval{H_{\rm targ}}{\gamma,\beta}$, в то время, как минимизация $E(\gamma,\beta)$ по параметрам производится с использованием классического компьютера. 
Однако, такой кванто-классический гибридный подход сталкивается с проблемами многопараметрической оптимизации и статистических ошибками при расчёте $E(\gamma,\beta)$.
В оригинальной работе \cite{Farhi2014} предлагался другой вариант алгоритма: в работе представлен эффективный классический метод вычисления $E(\gamma,\beta)$ для задачи Max-Cut на 3-регулярных графах, что снимает описанную выше проблему статистического шума измерений. Т.е. оптимальные или квази-оптимальные параметры $\{\beta_i, \gamma_i\}$ вычисляются полностью на классическом компьютере, квантовый же компьютер используется исключительно для сэмплирования результатов измерений состояния $\ket{\beta,\gamma}.$ Данный подход, помимо решения указанной выше проблемы позволяют получить некоторые аналитические оценки эффективности, но ограничивает класс возможных решаемых задач, т.к. необходима процедура эффективного классического вычисления $E(\gamma,\beta)$. 
В рамках работы над Дорожной картой использовался ещё один альтернативный вариант алгоритма из семейства QAOA, основанный на эмпирическом наблюдении близости квази-оптимальных параметров $\{\beta_i, \gamma_i\}$ для задач одного типа (например, Max-Cut). Впервые данная гипотеза была выдвинута в работе~\cite{brandao2018fixed} и получила дальнейшее развитие \cite{galda2021transferability, wurtz2021fixed}. Данная гипотеза говорит о возможности использовать одни и те же фиксированные параметры (углы) для различных задач, полностью исключая классическую оптимизацию вне процедуры поиска самих универсальных углов. В недавней работе \cite{shaydulin2024evidence} совмещение алгоритма QAOA с фиксированными углами с подходом алгоритма Гровера позволило наблюдать квантовое превосходство в задаче двоичных последовательностей с низкой автокорреляцией (low atocorrelation binary sequences, LABS).

В рамках текущей работы использовался оригинальный подход к поиску универсальных углов с использованием эвристической глобальной оптимизации на обучающей выборке задач~\cite{chernyavskiy2023fixed}. Пусть имеется некоторая обучающая выборка задач $C_i$ некоторого общего класса (например, Max-Cut). Для каждой задачи $C_i$, задаваемой соответствующим гамильтонианом задачи, и набора углов $(\gamma,\beta)$ имеется выходное состояние $\ket{\beta,\gamma}_i$ схемы QAOA. Тогда универсальные углы могут определяться оптимизацией некоторой интегральной метрики на наборе задач, например 
$$\max\limits_{\beta,\gamma} \min\limits_i P_C(\ket{\beta,\gamma}_i),$$ 
где $P_C(\ket{\beta,\gamma}_i)$ -- вероятность получения правильного ответа задачи $C_i$ при измерении состояния $\ket{\beta,\gamma}_i$. Далее будут описаны три алгоритма, построенные на основе алгоритма QAOA с фиксированными параметрами (fpQAOA). 

Так на основе fpQAOA был разработан \textbf{\textit{алгоритм из семейства алгоритмов решения системы линейных уравнений.}} Наиболее известным квантовым алгоритмом решения СЛАУ является HHL-алгоритм \cite{harrow2009quantum}, который, несмотря на гипотетическую возможность экспоненциального ускорения, имеет ряд серьезных ограничений и сложностей реализации \cite{scherer2017concrete, aaronson2015read}. Альтернативным подходом использования квантовых вычислителей (в основном, реализующих квантовый отжиг) является сведение задачи решения СЛАУ к QUBO \cite{jun2024qubo, lee2022effective}. В алгоритме, разработанном в рамках Дорожной карты, сведение системы линейных уравнений $Ax=B$ к QUBO осуществлялось переходом к минимизации квадратичного функционала 
\begin{equation}
x^T(A^TA)x + x^T(-2A^Tb)
\end{equation}
и бинаризацией переменных, далее для решения задачи применялся алгоритм fpQAOA. Найденные на тестовой выборке универсальные угла алгоритма fpQAOA показали наличие ускорения относительно случайного выбора, что говорит о работе гипотезы о близких углах для данного класса задач. Отдельно отметим, что разработанный алгоритм, несмотря на вероятностную структуру, ползволяет получить ответ с гарантированной точностью, а именно, с невязкой не выше $t\sqrt{n}2^{1-k},$ где $t$ -- верхняя оценка нормы матрицы $A$, $n$ -- размерность СЛАУ, а $k$ -- число битов бинаризации переменных.  

Схожий подход был использован для реализации \textbf{\textit{алгоритма решения дифференциальных уравнений.}} Задача решения обыкновенного дифференциального уравнения второго порядка с переменными коэффициентами 
\begin{equation}
f_2(t)y''(t)+f_1(t)y'(t)+f_0(t)y(t)+g(t)=0
\end{equation}
с граничными условиями первого рода может быть сведена к QUBO разностной аппроксимацией на сетке и бинаризацией переменных. Также, как и в случае с алгоритмом решения СЛАУ, гипотеза универсальности углов алгоритма fpQAOA подтвердилась на задачах QUBO, сформированных на верификационном классе дифференциальных уравнений вида $y''(t)=-g(1+kt^2)$. Отметим, что разработанный алгоритм требует достаточно большого числа кубитов, а именно $(n_t-2)k$, где $n_t$ -- число шагов сетки, $k$ -- число битов бинаризации переменных. Данный факт ограничивает точность вычислений на текущих квантовых процессорах и симуляторах. Для борьбы с данной проблемой алгоритм был усовершенствован итерационным подходом с уточнением решений -- на каждом новом шаге решение ищется как меньшая (и, соответственно, с большей точностью бинаризации) добавка к предыдущему. 

Ещё одним алгоритмом, разработанным на основе fpQAOA, является усовершенствованый алгоритм факторизации целых чисел на основе метода Шнорра \cite{zalivako2025shnorr}. В конце 2022 года был представлен препринт статьи \cite{yan2022factoring}, вызвавший большой резонанс в научном сообществе в связи с утверждением о возможности факторизации 2048-битного ключа криптографического протокола RSA с использованием всего лишь 372. Авторы работы предложили нестандартный метод использования алгоритма QAOA: алгоритм используется не для решения самостоятельной задачи, а для уточнения приближенного решения, полученного классическим алгоритмом. Метод факторизации Шнорра \cite{schnorr2021fast} сводит задачу к многократному поиску ближайших векторов на решетке, для чего в свою очередь используется классический LLL-алгоритм \cite{lenstra1982factoring}, результат работы которого как раз и предлагается уточнять с использованием квантового компьютера и алгоритма QAOA. Однако, в представленном авторами алгоритме имеются серьезные трудности, причем, как в классической, так и в квантовой части \cite{grebnev2023pitfalls, Khattar2023}. Отметим, однако, что возможность получения эффективного алгоритма факторизации на основе такого подхода не опровергнута \cite{zalivako2025shnorr}. 

Использование же фиксированных углов вместо квантово-классической оптимизации позволяет существенно повысить эффективность работы алгоритма. Как и в случае с алгоритмами решения СЛАУ и дифференциальных уравнений, гипотеза близких оптимальных углов подтвердилась.

\subsection{Алгоритмы квантовой химии}
Задача поиска спектра собственных значений линейных эрмитовых операторов, соответствующих наблюдаемым физическим величинам, является одной из важнейших задач квантовой физики.
В частности, расчёт энергии основного состояния --- минимального собственного значения гамильтониана физической системы --- это ключ к получению множества других производных физических величин.
Потенциально эффективным подходом к решению данной задачи являются квантовые вариационные алгоритмы --- гибридные квантово-классические методы, впервые предложенные в контексте поиска энергии основного состояния иона гидрида гелия с использованием фотонного квантового процессора~\cite{vqe_first}.
Они позволили найти компромисс: использовать большое количество квантовых цепочек малой глубины вместо одной глубокой квантовой цепочки, применявшейся в ранее предложенных квантовых алгоритмах, решавших аналогичную задачу~\cite{Kitaev1995QuantumMA, aspuru2005simulated}.

Исходно разработанным алгоритмом, лежащим в основе остальных вариационных квантовых алгоритмов, был алгоритм \textbf{\textit{алгоритм квантового вариационного поиска собственных значений}}. В подобных алгоритмах параметризованное квантовое состояние --- анзац $|\psi(\boldsymbol{\theta})\rangle$ --- создаётся с помощью квантовой цепочки, зависящей от набора параметров $\boldsymbol{\theta}$. 
Затем выполняется измерение среднего значения гамильтониана \( H \) на этом состоянии:
\begin{equation}
E(\boldsymbol{\theta}) = \frac{\langle \psi(\boldsymbol{\theta}) | H | \psi(\boldsymbol{\theta}) \rangle}{\langle \psi(\boldsymbol{\theta}) | \psi(\boldsymbol{\theta}) \rangle},
\end{equation}
после чего параметры $ \boldsymbol{\theta} $ оптимизируются выбранным классическим алгоритмом минимизации. Соответственно, выполняется верхняя оценка минимального собственного значения гамильтониана $H$: $E_0 = \min_{\boldsymbol{\theta}} E(\boldsymbol{\theta})$.

На следующем этапе был реализован \textit{\textbf{алгоритм поиска основного состояния молекулы}}. В данном алгоритме используется гамильтониан электронной системы молекулы в приближении Борна-Оппенгеймера: 
\begin{equation}\label{fermionic_ham}
    H = \sum_{pq} h_{pq} a_p^\dagger a_q + \frac{1}{2} \sum_{pqrs} h_{pqrs} a_p^\dagger a_q^\dagger a_r a_s,
\end{equation}
где $a_p^\dagger$ и $a_q$ --- фермионные операторы рождения и уничтожения, $h_{pq}$, $h_{pqrs}$ --- одноэлектронные интегралы и двухэлектронные интегралы соответственно.
В качестве анзаца в данном алгоритме используется Unitary Coupled Cluster (UCC) анзац~\cite{romero2018strategies}, в нём квантовое состояние параметризуется как $|\psi(\boldsymbol{\theta})\rangle = e^{T(\boldsymbol{\theta}) - T^\dagger(\boldsymbol{\theta})} |\psi_0\rangle$, где $T(\boldsymbol{\theta})$ --- кластерный оператор возбуждения, а $|\psi_0\rangle$ --- состояние Хартри-Фока. Отличительной особенностью текущей имплементации данного алгоритма является то, что в неё встроена автоматическая группировка слагаемых гамильтониана на основе свойств их взаимной коммутации. 
Это позволяет сократить количество необходимых алгоритму квантовых цепочек, а следовательно и количество необходимых измерений, которые нужно выполнить на квантовом устройстве. 
Разновидность этого алгоритма использовалась для моделирования энергетических характеристик реакции окисления угарного газа~\cite{sapova2022variational}.

\begin{figure*}
    \centering
    \includegraphics[width=0.9\textwidth]{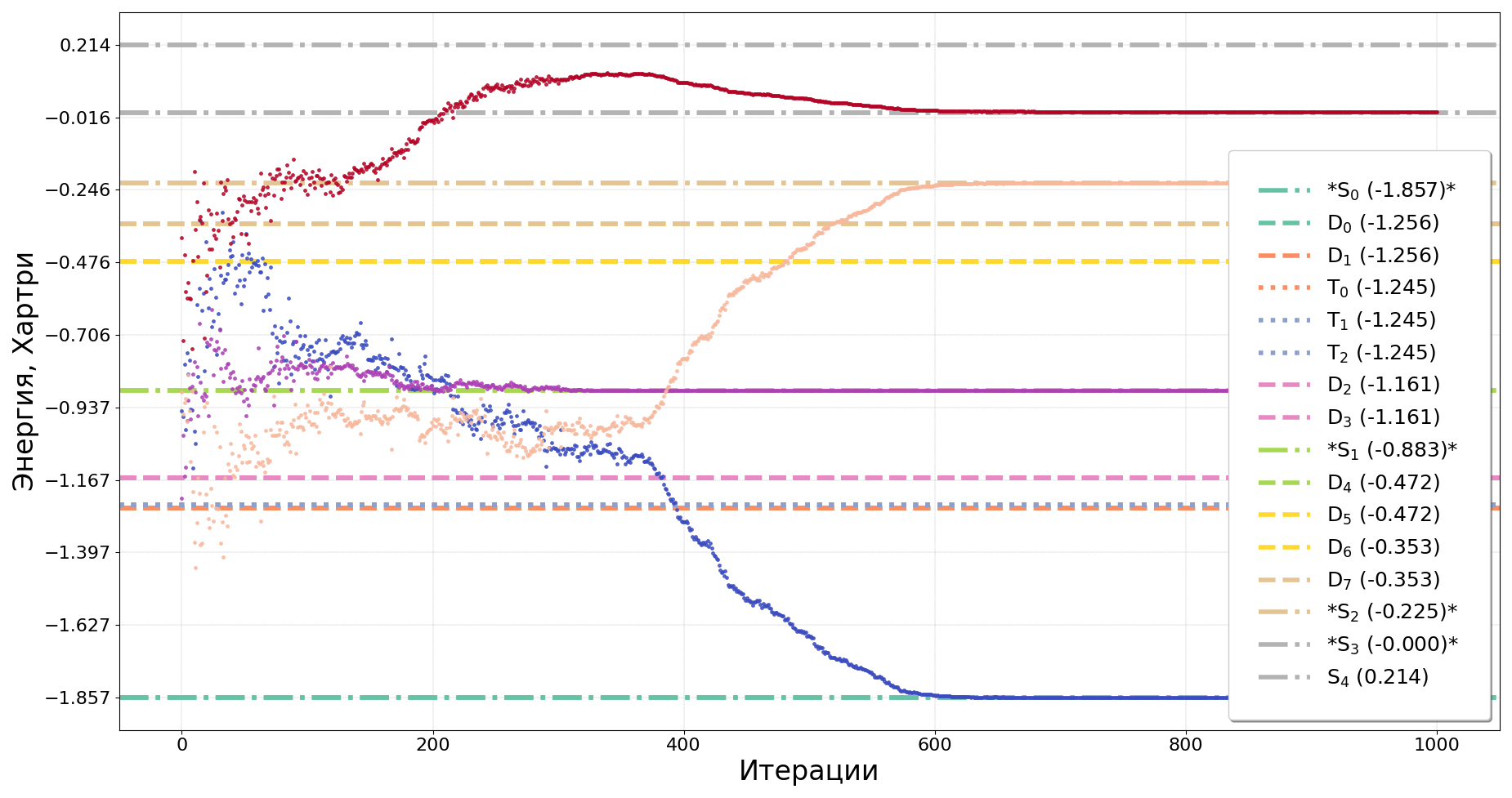}
    \caption{Результаты, полученные с помощью алгоритма моделирования спектров возбуждения молекулы. Расчёт произведён для 4 синглетных состояний молекулы H$_2$ в базисе STO-3G. Найденные алгоритмом синглетные энергии помечены символом *.}
    \label{fig:excited}
\end{figure*}

Последним из группы вариационных квантовых алгоритмов был разработан \textit{\textbf{алгоритм моделирования спектров возбуждения молекулы}}.
В данном алгоритме целевая функция, подлежащая оптимизации, выглядит иначе: 
\begin{multline}\label{cost_function}
    L_\omega(\boldsymbol{\theta}) = \sum_{j=0}^{k} \omega_j \langle \varphi_j | U^\dagger(\boldsymbol{\theta}) |H| U(\boldsymbol{\theta}) | \varphi_j \rangle \\
    +p \langle \varphi_j | U^\dagger(\boldsymbol{\theta}) |S^2| U(\boldsymbol{\theta}) | \varphi_j \rangle,
\end{multline}
здесь $\omega_j$ --- набор весовых коэффициентов для начальных ортогональных состояний Хартри-Фока, $U(\boldsymbol{\theta})$ --- оператор, символизирующий анзац, $p$ --- весовой коэффициент для ограничения, $S^2$ --- оператор квадрата полного спина.
Алгоритм одновременно ищет $k$ состояний из спектра собственных состояний молекулы, сохраняя их исходно заданную ортогональность. 
Особенностью реализации алгоритма является возможность находить синглетные состояния молекулы за счёт дополнительного слагаемого --- ограничения в целевой функции (\ref{cost_function}), фиксирующего собственное значение оператора полного спина молекулы. 
На Рис.~\ref{fig:excited} представлен расчёт возбуждённых состояний, выполненный для молекулы H$_2$.
Показан процесс минимизации энергии для четырёх синглетных состояний молекулы, цветные горизонтальные линии символизируют энергии, полученные с помощью классического метода Full Configuration Interaction (FCI). Синглетные состояния обозначены как S$_n$, те, что были найдены в ходе оптимизации, дополнительно помечены символом *. Для расчётов использовался минимальный базис --- STO-3G.

Все вышеупомянутые квантовые вариационные алгоритмы способны находить значения энергий с точностью, сопоставимой с методом FCI в пределах химической точности --- $0.001$ Хартри относительно выбранного базиса, что соответствует результатам, полученным с помощью других квантовых вычислительных устройств и их эмуляторов~\cite{PhysRevResearch.5.033071,bench}.
Следующим шагом может быть разработка алгоритмов с адаптивным анзацем, который динамически изменяется в процессе вычислений, а также алгоритмов, не требующих постоянного обмена данными между квантовой и классической частями.
Кроме того, перспективным направлением является разработка более совершенных методов оптимизации количества требуемых алгоритму квантовых цепочек, оптимизации глубины квантовых цепочек и числа двухкубитных операций.

Помимо вариационных квантовых алгоритмов, задача поиска энергии основного состояния электронной системы молекулы может быть решена альтернативным способом --- сведением её к задаче QUBO. Такой подход позволяет представить исходный гамильтониан в виде квадратичной формы с бинарными переменными, эквивалентной гамильтониану Изинга, что делает задачу удобной для решения методами квантового отжига~\cite{xia2017electronic, chermoshentsev2021polynomial}:
\begin{equation}\label{qubo}
H_{\text{QUBO}} = \sum_{i} a_i x_i + \sum_{i < j} b_{ij} x_i x_j, \quad x_i \in \{0,1\},
\end{equation}
где $x_i$ --- бинарные переменные, $a_i$ --- линейные коэффициенты, $b_i$ --- коэффициенты взаимодействия.
Разработанный \textit{\textbf{алгоритм квантовой химии на основе решения
задачи Изинга}} выполняет квадратизацию исходного гамильтониана (\ref{fermionic_ham}) и находит вектор основного состояния, используя алгоритм {SimBif}. Алгоритм способен находить энергию основного состояния двухатомных молекул H$_2$, He$_2$, HeH$^{+}$ с точностью, которая сравнима с методом FCI в пределах $0.01$ Хартри для минимального базиса STO-3G. Однако для дальнейшего развития данного подхода потребуется разработка более совершенных методов приведения операторов гамильтониана (\ref{fermionic_ham}) к форме (\ref{qubo}), которые позволят избежать быстрого роста требуемых вычислительных ресурсов с увеличением размера электронной системы молекулы.

\subsection{Алгоритмы моделирования физических систем}

Квантовые алгоритмы симуляции квантовой динамики играют ключевую роль в исследованиях квантовых систем, квантовой химии и материаловедения. Они позволяют эффективно моделировать эволюцию квантовых систем. Основной задачей таких алгоритмов является приближённое вычисление унитарного оператора эволюции $U(t) = \exp\left(-\imath Ht\right)$, где $H$ – гамильтониан системы, $t$ – время эволюции. Основным подходом к построению квантовых цепочек приближающих оператор $U(t)$ на квантовом компьютере является \textit{разложение Сузуки-Троттера}~\cite{Berry2007, Hatano2005, PhysRevResearch.5.023199, Dhand_2014, PhysRevX.11.011020}, реализованное в рамках \textit{\textbf{алгоритма моделирования магнитных материалов}}. Идея разложения Сузуки-Троттера состоит в том, чтобы выразить гамильтониан как сумму легко моделируемых гамильтонианов, а затем аппроксимировать общую эволюцию как последовательность более простых эволюций. Допустим имеется представление моделируемого гамильтониана $H$ в виде суммы $H = \sum_{i=1}^{N}H_{i}$, тогда, разложение оператора эволюции данного гамильтониана может быть записано в виде 
\begin{equation}\label{eq:trotterization}
    U\left(t\right)=e^{-it\sum_{j=1}^{N}H_j}=\prod_{j=1}^Ne^{-itH_j}+O\left(N^2t^2\right).
\end{equation}
В случае если время необходимой симуляции достаточно мало $t \ll 1$ ошибкой аппроксимации можно пренебречь. При этом, данное приближение становится точным в том случае, если члены гамильтониана взаимно коммутируют $\left[H_i,H_j\right]=0,~i \neq j$. Следовательно, все ошибки в приближении проистекают из взаимной некоммутативности членов гамильтониана. 

В том случае, если необходимое время симуляции системы достаточно большое, формулу~\eqref{eq:trotterization} можно преобразовать. Разбивая временной интервал симуляции $t$ на $r$ одинаковых по длительности меньших временных интервалов, можно получить следующую формулу 
\begin{equation}
    U\left(t\right)=\left(\prod_{j=1}^Ne^{-iH_jt/r}\right)^{r}+O\left(N^2t^2/r\right).
\end{equation}
Так можно получить произвольную точность аппроксимации унитарной динамики системы заложив в симуляцию необходимое число шагов. 

Более точные аппроксимации можно построить, составив последовательность операторных экспонент, исключающую ошибки. Такие схемы называются разложениями Сузуки-Троттера высших порядков. Для второго порядка ($k=2$) формула Сузуки-Троттера имеет вид
\begin{equation}
    U_{2\left(t\right)}=\left(\prod_{j=1}^Ne^{-iH_jt/2r}\prod_{j=N}^1e^{-iH_jt/2r}\right)^r+ O\left(N^3t^3/r^2\right).
\end{equation}
Формулы порядков $k>2$ могут быть выражены рекурсивно через формулу второго порядка как
\begin{equation}
\begin{aligned}
&U_{2k}\left(t\right)= \\
&= \left[U_{2k-2}\left(s_kt\right)\right]^2U_{2k-2}\left(c_{k}t\right)\left[U_{2k-2}\left(s_kt\right)\right]^2+\\
&+O\left(\left(Nt\right)^{2k+1}/r^{2k}\right),    
\end{aligned}
\end{equation}
где $s_k=\left(4-4^{1/\left(k-1\right)}\right)^{-1}$. При этом, все высшие порядки могут быть только четными. Так как ошибка аппроксимации в формуле Сузуки-Троттера появляется ввиду некоммутативности членов гамильтониана, имеет смысл представить гамильтониан в виде суммы соответствующих семейств так, чтобы внутри семейства все члены взаимно коммутировали, в то время как сами семейства могут некоммутировать. Такой подход позволяет уменьшить глубину цепочки, но при этом в общем случае задача поиска разбиения членов гамильтониана на семейства эквивалентна нахождению раскраски графа и является NP-трудной~\cite{garey1974some}. Однако для большинства практических приложений решение задачи в общем виде не требуется, либо достаточно ограничиться эвристикой~\cite{e21121218}. Текущая имплементация алгоритма опирается на эвристический подход для группировки членов локальных гамильтонианов (типичных для магнитных материалов), что дает существенное сокращение глубины квантовой цепочки.

Альтернативный подход к моделированию квантовой динамики реализован в качестве \textit{\textbf{алгоритма моделирования квантвой динамики многочастичных систем}}. Данный алгоритм реализует симуляцию гамильтониана на основе приема известного как квантовая обработка сигнала (quantum signal processing, QSP)~\cite{martyn_grand_2021, dong_efficient_2021, rossi_multivariable_2022, skelton_mostly_2024, ying_stable_2022, ying_stable_2022}. Суть QSP заключается в принципиальной возможности реализации полиномиальных преобразований подматриц специальных унитарных операторов, называемых блочным кодированием (block encoding, BE)~\cite{sunderhauf_block-encoding_2024, camps_fable_2022, lamm_block_2024, boyd_low-overhead_2024}. Если заданный гамильтониан представить в виде BE, при помощи QSP возможно сконструировать такую квантовую цепочку, что ее действие на состояние входного регистра будет соответствовать действию оператора $P_d\left(H\right)$, где $P_d$ – полином степени $d$. В таком случае, приближая комплексную экспоненту полиномом степени $d$ можно вычислить матричную экспоненту от гамильтониана, получая оператор эволюции~\cite{gilyen_qsvt, low_optimal_2017, martyn_efficient_2023}.

В общем случае задача построения BE для заданного гамильтониана допускает множество решений. В рассматриваемом случае вводится ограничение на представление гамильтониана в виде линейной комбинации (linear combination of unitaries, LCU) строк Паули с соответствующими коэффициентами
\begin{equation}\label{eq:hamiltonian}
    H=\sum_{i=1}^{N}{h_i{P}_i}.
\end{equation}
В таком случае, BE гамильтониана может быть представлено как последовательность двух преобразований: $\texttt{PREPARE}$ и $\texttt{SELECT}$. 

Операция $\texttt{PREPARE}$ призвана записать в квантовый регистр кубит-анцилл коэффициенты разложения гамильтониана (матрицы) на строки Паули. Действие оператора $\texttt{PREPARE}$ можно описать следующим образом
\begin{equation}
    \ket{h} = \texttt{PREPARE}\ket{0}^{\otimes N} = \sum_{i=1}^{N} \sqrt{\frac{h_{k}}{||h_{k}||}}\ket{i},
\end{equation}
где $||h|| = \sum_{k}|h_{k}|$. Преобразование $\texttt{PREPARE}$ приготавливает состояние суперпозиции, в котором число кубит соответствует $\lceil\log_{2}(N)\rceil$, а амплитуды пропорциональны самим коэффициентам разложения гамильтониана по строкам Паули. 

Операция $\texttt{SELECT}$ осуществляет непосредственно преобразования состояния во входном регистре, т.е. осуществляет преобразование вида $\ket{\psi} \rightarrow H\ket{\psi}$, где $H$ – гамильтониан. Имея дополнительный регистр кубит-анцилл в состоянии $\texttt{PREPARE}\ket{0}^{\otimes N}$ и действуя с этого регистра условными операторами ${P}_k$ можно получить эффективное действие оператора $\sum_{k}{h_k{P}_k}$ на состояние во входном регистре. Операция $\texttt{SELECT}$ действует следующим образом
\begin{equation}
    \texttt{SELECT}\ket{h}\ket{\psi} = \sum_{i=1}^{N} \sqrt{\frac{h_{k}}{||h_{k}||}}\ket{i}P_{k}\ket{\psi},
\end{equation}
Комбинация преобразований $\texttt{PREPARE}$ и $\texttt{SELECT}$ задает BE гамильтониана вида~\eqref{eq:hamiltonian}. При этом, имеющееся унитарное преобразование $U_{\rm BE}$ имеет блочную структуру
\begin{equation}
    U_{\rm BE} =\left[\begin{matrix}H&\ast\\\ast&\ast\\\end{matrix}\right], 
\end{equation}
а требующееся преобразование над состоянием входного регистра вида $\ket{\psi} \rightarrow H\ket{\psi}$ будет иметь место только в том случае, если при измерении регистра кубит-анцилл результат будет равен $0\ldots0$. 

Имея BE заданного гамильтониана возможным является получение внутри квантового алгоритма преобразования вида $P_d\left(H\right)$. Прием QSP по сути является унитарным анзацем вида
\begin{equation}\label{eq:qsvt}
    U_\Phi=e^{-i\phi_0Z_\Pi}\prod_{j=1}^{d}{\left(U_{\rm BE}e^{-i\phi_kZ_\Pi}\right)\ =\ \left[\begin{matrix}P_d\left(H\right)&\ast\\\ast&\ast\\\end{matrix}\right]},
\end{equation}
где ${(\phi}_0, \phi_1,...,\phi_d)$ – фазовые параметры. Из выражения~\eqref{eq:qsvt} можно видеть, что оператор $U_\Phi$ является ничем иным как BE оператора $P_d\left(H\right)$. 

При помощи BE заданного гамильтониана и метода QSP можно получить полиномиальное преобразование гамильтониана в квантовом компьютере. Конкретный вид полинома задается фазовыми параметрами ${(\phi}_0,\phi_1,...,\phi_d)$. Однако, для нахождения оператора эволюции требуется вычисление экспоненты от заданного гамильтониана. Разумным в такой ситуации является представление экспоненты в виде полинома при помощи приближения Якоби-Ангера. Оператор эволюции можно представить в виде $U(t) =\cos(tH)-i\sin(tH)$. Приближая полиномами оба члена, получим 
\begin{equation}
\begin{aligned}
&\sin{\left(tx\right)}=2\sum_{k=0}^{\infty}\left(-1\right)^kJ_{2k+1}\left(t\right)T_{2k+1}\left(x\right),\\
&\cos{\left(tx\right)}=J_0\left(t\right)+2\sum_{k=1}^{\infty}\left(-1\right)^kJ_{2k}\left(t\right)T_{2k}\left(x\right),    
\end{aligned}
\end{equation}
где $J_k$ – функции Бесселя первого порядка, $T_k$ – полиномы Чебышева первого порядка. Для приближения $\sin{\left(tx\right)}$ и $\cos{\left(tx\right)}$ с точностью $\varepsilon/\sqrt{\left(2\right)}$  требуется 
\begin{equation}
r=\Theta\left(t+\frac{\log{\left(1/\varepsilon\right)}}{\log{\left(e+\log{\left(1/\varepsilon\right)}/t\right)}}\right)
\end{equation}
членов соответствующего ряда. Такой скейлинг дает оценку на сложность симуляции гамильтониана. Однако, для моделирования динамики на практике, требуется найти фазовые параметры ${(\phi}_0,\phi_1,...,\phi_d)$ приближающие функции $\sin{\left(tx\right)}$ и $\cos{\left(tx\right)}$~\cite{dong_efficient_2021, skelton_mostly_2024, ying_stable_2022}. Для решения этой задачи алгоритм использует методы оптимизации типа методов деформируемого многогранника (алгоритм Нелдера-Мида).

Отдельное место занимают подходы, позволяющие осуществить \emph{\textbf 
 {сверхвременное моделирование}} квантовой динамики интересующей системы. Посредством сверхвременного моделирования динамика определенных квантовых систем может быть смоделирована с сублинейной по времени эволюции~\cite{gu_fast-forwarding_2021, atia_fast-forwarding_2017, kokcu_fixed_2022}. Результат~\cite{atia_fast-forwarding_2017} говорит о невозможности сверхвременного моделирования для произвольного гамильтониана. Однако это возможно для гамильтонианов систем, принадлежащих определенному классу. К таким гамильтонианам относятся, например, модель Изинга в поперечном поле (TFIM) и XY-модель в поперечном поле (TFXY)~\cite{kokcu_fixed_2022}. Конструктивный алгоритм на основе разложения Картана алгебры Ли, генерируемой гамильтонианом~\cite{kokcu_fixed_2022} позволяет для систем подобного рода построить квантовую цепочку, моделирующую квантовую динамику с глубиной, независимой от времени. Данный подход был реализован для экспериментальной симуляции динамики TFIM на ионном квантовом процессоре, разрабатываемом в  ФИАН им. П. Н. Лебедева \cite{zalivako2025ufn}.

\section{Мониторинг и бенчмаркинг}\label{moniben}

Модуль мониторинга и бенчмаркинга позволяет отслеживать состояние квантового вычислителя на различных уровнях. Мониторинг дает базовое представление о состоянии квантового устройства: количество кубитов, скорости операций, точность измерений, а также точность одно- и двухкубитных операций. Данные показатели измеряются поставщиком квантового устройства.

Бенчмаркинг представляет собой набор методов, которые анализируют качество и производительность квантового процессора \textbf{на логическом уровне}, вне зависимости от его физической реализации. Базовым методом бенчмаркинга является процедура квантовой томографии. И хотя в своей основной формулировке она плохо масштабируется на большое число кубитов, квантовая томография позволяет с высокой достоверностью анализировать различные элементы квантового вычислителя, включая процедуры приготовления, преобразования и измерения.

Разные виды квантовой томографии анализируют различные объекты. В \textit{томографии квантовых состояний} (Quantum State Tomography, QST) изучаемым объектом является матрица плотности $\rho$, описывающая состояние кубита или системы кубитов (в том числе запутанных) \cite{bantysh2021quantum}. Основные свойства матрицы плотности: неотрицательная определенность ($\rho \geq 0$) и нормированность ($\Tr\rho = 1$).

\textit{Томография квантового процесса} (Quantum Process Tomography, QPT) направлена на полную характеристику линейного преобразования $\mathcal{E}(\rho)$, которое действует на произвольное входное состояние \cite{mohseni2008quantum}. Важными свойствами такого преобразования являются полная положительность (для любого $\rho$ действие канала на подсистему приводит к неотрицательной матрице, т.е. $(\mathcal{I}\otimes\mathcal{E})(\rho) \geq 0$) и сохранение следа (для любого $\rho$ выполняется $\Tr(\mathcal{E}(\rho)) = \Tr(\rho)$). Эти условия составляют так называемое CPTP-условие (Complete Positive Trace Preserving). Физическое описание квантового канала может быть представлено различными способами, и для каждого из них CPTP-условие принимает разную форму. Одним из наиболее распространённых представлений является описание через матрицу Чоя $\chi$ размерности $d^2 \times d^2$: $\mathcal{E}(\rho) = \sum_{mn} \chi_{mn} A_m \rho A_n^\dagger$, где матрицы ${A_m, m=1,\dots,d^2}$ образуют базис в пространстве матриц размерности $d \times d$. В этом представлении условие полной положительности эквивалентно неотрицательности матрицы $\chi$, а условие сохранения следа выражается как $\sum_{mn} \chi_{mn} A_n^\dagger A_m = I_d$, где $I_d$ — единичная матрица размерности $d \times d$. Стандартный базис вида
\begin{equation}
    A_1 = \mqty(1&0&\cdots\\0&0&\cdots\\\vdots&\vdots&\ddots), \quad
    A_2 = \mqty(0&1&\cdots\\0&0&\cdots\\\vdots&\vdots&\ddots), \quad
    \dots
\end{equation}
соответствует условию взятия частичного следа по подсистеме $A$, если представить, что матрица $\chi$ соответствует системе $AB$, где каждая подсистема имеет размерность $d$: $\Tr_A(\chi)=I_d$.

\textit{Томография квантового детектора} (Quantum Detector Tomography, QDT) направлена на определение POVM-эффектов (positive operator-valued measure), описывающих квантовый детектор \cite{chen2019detector}. Эти эффекты задаются набором неотрицательных матриц $E_k$, где каждая матрица соответствует конкретному исходу измерения, а вероятность этого исхода для входного состояния $\rho$ определяется как $p_k = \Tr(E_k \rho)$. Условие нормировки вероятностей задает ограничение на POVM-эффекты: $\sum_k{E_k} = I_d$.

Квантовая томография включает два ключевых этапа: выполнение информационно полного набора квантовых измерений, который содержит всю необходимую информацию о параметрах исследуемого объекта, и реконструкцию этих параметров с учётом статистических флуктуаций измерений. На этапе выполнения измерений модуль использует стандартные факторизованные измерения Паули, где каждый кубит измеряется в трёх базисах, соответствующих операторам Паули: $\sigma_x$, $\sigma_y$ и $\sigma_z$. В томографии квантового процесса и квантового детектора применяется стандартный набор факторизуемых состояний, в котором каждый кубит подготавливается в одном из четырёх состояний ${\ket{0}, \ket{1}, \ket{+}, \ket{+i}:=2^{-1/2}(\ket{0} + i\ket{1})}$, образующих полный базис в пространстве матриц плотности.

После получения результатов измерений осуществляется процедура реконструкции параметров квантового состояния, процесса или детектора. Важно отметить, что все рассматриваемые объекты линейно связаны с вероятностями обнаружения различных событий, что позволяет решать задачу нахождения неизвестных параметров через систему линейных алгебраических уравнений. Однако, результат реконструкции может не удовлетворять физическим ограничениям, и в таком случае выполняется проецирование на множество корректных решений.

Этот метод используется как начальное приближение для более точного метода максимального правдоподобия (ММП). В основе ММП лежит максимизация функции логарифмического правдоподобия: 
\begin{equation}
    L(\vec{\theta}) = \sum_{\alpha,i}{k_{\alpha,i}\ln{p_{\alpha,i}(\vec{\theta})}},
\end{equation}
зависящей от неизвестных параметров $\vec{\theta}$. Здесь индекс $\alpha$ пробегает по всему набору томографических схем, $i$ -- индекс результата измерения, $k_{\alpha,i}$ -- число наблюдений соответствующего результата, $p_{\alpha,i}(\vec{\theta})$ -- модельная вероятность данного результата, зависящая от искомых параметров. Поскольку на все рассматриваемые объекты накладывается условие неотрицательности ($\rho\geq0$ для состояний, $\chi\geq0$ для процессов, $E_k\geq0$ для POVM-эффектов), в модуле используется корневая параметризация \cite{bogdanov2004statistical,bogdanov2009unified}. В этом случае вместо поиска неотрицательной матрицы $X$ размерности $d \times d$ ищется комплексная матрица $c$ размерности $d \times r$, а матрица $X$ представляется в виде $X = cc^\dagger$. Параметр $r$ является гиперпараметром алгоритма и позволяет фиксировать ранг матрицы $X$. Условие нормировки накладывается с помощью метода множителей Лагранжа \cite{bogdanov2009unified} или применяя проективный градиентный спуск (в модуле использовался алгоритм из работы \cite{li2015accelerated}).

В частности, значение ранга $r=1$ соответствует случаю чистых состояний, унитарных процессов и проекционных измерений. Единичный ранг также используется в задаче \textit{томографии Гамильтониана} (Quantum Hamiltonian Tomography, QHT). После реконструкции матрицы $\chi$ вычисляется её собственный вектор $\vec{e}$, соответствующий максимальному собственному значению. В этом случае унитарная матрица процесса $U$ определяется как $U = \sqrt{d} \sum_{m} e_m A_m$, а эффективный Гамильтониан вычисляется по формуле $H_{\rm eff} = i \ln(U) / \tau$, где $\tau$ — время выполнения операции.

Описанные методы квантовой томографии применялись к анализу гейта Мёльмера--Соренсона (MS) ионного квантового вычислителя. Для реализации процедур приготовления и измерения использовались нативные операции
\begin{equation}
R_\varphi(\theta)=\exp[-i\frac{\theta}{2}(\sigma_x\cos\varphi+\sigma_y\sin\varphi)],
\end{equation}
реализующие вращение кубита на сфере Блоха вокруг экваториальной оси, лежащей под углом $\varphi$ к оси $x$, на угол $\theta$.

В рамках развития аппарата квантовой томографии был также проработан метод получения доверительных интервалов на аффинные функции от восстанавливаемых состояний (матриц плотности $\rho$) и квантовых каналов (соответствующих матриц Чоя $\rho_{\rm choi}$)~\cite{kiktenko2021confidence, norkin2024reliable}.
К важным примерам таких функций относится фиделити восстанавливаемого состояния $\rho$, по отношению к чистому целевому состоянию $\ket{\psi_{\rm targ}}$:
\begin{equation}
    F(\rho)=\bra{\psi_{\rm targ}}\rho\ket{\psi_{\rm targ}},
\end{equation}
и фиделити запутанности восстанавливаемого канала $\rho_{\rm Choi}$ по отношению к целевому унитарному преобразованию $U_{\rm targ}$ :
\begin{equation}
    F(\rho_{\rm Choi})=\bra{\Phi}(U^\dagger_{\rm targ}\otimes I) \rho_{\rm Choi} (U_{\rm targ}\otimes I)\ket{\Phi},
\end{equation}
где $\ket{\Phi}=d^{-1/2}\sum_i\ket{i}\otimes\ket{i}$ обозначает максимально запутанное состояние двух кудитов размерности $d$.
Разработанные методы дают возможность получать семейство величин $F_{\rm min}, F_{\rm max}, {\rm CL}$, соответствующих доверительным интервалам вида
\begin{equation}
    \Pr[F_{\rm min}\leq F \leq F_{\rm max}] \geq {\rm CL},
\end{equation}
где $F$ -- истинное значение аффинной функции и вероятность рассматривается по отношению к получаемым экспериментальным результатам (``сэмплам'') для произвольного квантового состояния или канала.
Данный инструмент дает дополнительные полезные возможности по характеризации квантовых процессоров.

Квантовая томография позволяет получать подробное описание квантовых состояний, детекторов и каналов, однако зачастую интерес представляет определение только точности квантового канала. Для этого в модуле используется процедура \textit{рандомизированного бенчмаркинга} (Randomized Benchmarking, RB) \cite{magesan2012efficient}. В основе неё лежит ряд теорем квантовой информатики, согласно которым случайные квантовые схемы, состоящие из $L$ зашумлённых преобразований группы Клиффорда, могут быть аппроксимированы последовательным действием $L$ деполяризующих каналов вида $\mathcal{E}_\gamma(\rho)=\gamma\rho + \Tr(\rho)(1-\gamma) I_d/d$. Степень поляризации $\gamma$ при этом связана со средней точностью одного гейта Клиффорда по формуле $F=1/d+(1-1/d)\gamma$. Выполняя измерения для схем различной глубины $L$ можно получить зависимость вероятности корректного результата $p$ от $L$, которая должна иметь экспоненциальный вид $p(L)=A\gamma^L+B$. Это позволяет экспериментально оценить $\gamma$ и, как результат, среднюю точность $F$ клиффордовской операции. Ошибки, связанные с инициализацией и считыванием, при этом определяют параметры $A$ и $B$ и не влияют на величину $\gamma$.

Приведённые выше методы бенчмаркинга призваны оценивать отдельные составляющие элементы квантового процессора. Для оценки его возможностей как целого в модуле используется подход на основе квантового объёма (Quantum Volume, QV) \cite{cross2019validating}. Квантовый объём связан с задачей определения тяжелых выходов случайных квантовых схем специального вида. Каждая такая схема содержит $r\in\{1,2,3,\ldots,n_{\rm av}\}$ слоёв (где $n_{\rm av}$ -- максимально доступное количество кубитов в процессоре), в каждом из которых над случайными парами кубитов выполняются случайные двухкубитные преобразования.

Тяжелыми выходами такой квантовой схемы являются те битовые строки, вероятность возникновения которых выше медианной вероятности среди всех $2^n$ значений вероятностей. Статистически оценка гарантированной доли тяжелых выходов задаётся выражением
\begin{equation}
h_{\rm est} = \frac{n_h - \sqrt{n_h(n_s-n_h/n_c)}}{n_cn_s},
\end{equation}
где $n_h$ -- число обнаруженных тяжелых выходов, $n_c$ -- число случайных квантовых схем, $n_s$ -- число повторений каждой схемы. Заметим, что для расчёта величины $h_{\rm est}$ необходимо для каждой схемы рассчитать полный набор вероятностей, используя эмулятор идеального квантового процессора. Тест на определение доли тяжелых выходов считается пройденным при выполнении условия $h_{\rm est}>2/3$. Квантовый объём вычисляется по формуле $\text{QV}=2^{n_{\rm max}}$, где $n_{\rm max}$ -- максимальное число кубитов, при котором для всех $n\leq n_{\rm max}$ тест на определение доли тяжелых выходов в случайных $n$-кубитных схемах глубины $r=n$ успешно проходится.

Квантовый объём является высокоуровневой метрикой, которая затрагивает все уровни архитектуры квантового процессора, включая: связность кубитов, точность приготовления, преобразования и измерения состояния кубита, перекрёстное взаимодействие между кубитами, возможность параллельного выполнения операций.

Кроме того модуль мониторинга и бенчмаркинга включает в себя \emph{систему непрерывного мониторинга квантовых процессоров}~\cite{Zolotarev2023}.
Данный инструмент позволяет получать оценки на квантовые каналы (CPTP отображения), соответствующие зашумленной реализации гейтов на реальных квантовых процессорах, на основе анализа результирующих битовых строк, полученных по результатам запуска квантовых цепочек на этих процессорах.
В основе своего функционирования закладываются следующие модельные предположения о действии шумов: (i) каждому определенному однокубитному (двухкубитному) гейту на определенном кубите (паре кубитов) соответствует свой квантовый канал; (ii) вид квантовых каналов остается постоянным внутри цепочки, а также среди всех цепочек, поступивших на вход; (iii) зашумленному приготовлению или измерению соответствует действие некоторого квантового канала следующего за или предшествующего идеальному приготовлению или идеальному проективному измерению соответственно. 
В качестве целевой функции оптимизации используется логарифмическое правдоподобие, а в качестве основного инструмента оптимизации выступает аппарат оптимизации на Римановых многообразиях~\cite{pechen2008control, oza2009optimization, luchnikov2021riemannian}, реализованный в пакете QGOpt~\cite{luchnikov2021qgopt}.

\iffalse
\begin{figure}
    \centering
    %\includegraphics[width=0.5\linewidth]{}
    \caption{Схема функционирования системы непрерывного мониторинга~\cite{zolotarev2023continuous}.
    На вход поступает совокупность квантовых схем, запущенных на квантовом процессоре, и результирующие битовые строки, полученные по результатам реализации данных схем.
    На выходе выдаются оценки на квантовые каналы, соответствующие зашумленной реализации квантовых гейтов.
    }
    \label{fig:monitoring}
\end{figure}
\fi

\section{Методы компиляции и оптимизации} \label{sec:Compilation_and_optimization}

Для запуска квантового алгоритма на реальном физическом устройстве требуется преобразование квантовой цепочки в форму, содержащую операции, которые являются выполнимыми на конкретном физическом процессоре. 
Квантовые операции, напрямую выполняемые на выбранном процессоре, называют {\it нативными гейтами}, а
процедуры преобразования квантовой цепочки называют {\it транспиляцией} и {\it компиляцией}. 
В процессе транспиляции происходит преобразование последовательности гейтов в эквивалентную последовательность других гейтов, например, из одного универсального набора операций в другой. 
Компиляцией же называется преобразование квантовой цепочки в эквивалентную последовательность нативных гейтов для данной платформы и конкретного процессора. 
Для решения данных задач могут быть задействованы стандартные инструменты, например, из библиотек Qiskit \cite{qiskit2024} и Cirq \cite{cirq2024}. В дополнение к этому актуальной задачей является оптимизация данной цепочки – запуск её с использованием минимального количества операций, что помогает сэкономить ресурс квантовых процессоров. 

Для работы с кубитными процессорами, оперирующими двухуровневыми квантовыми системами, были разработаны методы компиляции и оптимизации. Изначальная цепочка преобразуется в эквивалентную последователь однокубитных и двухкубитных гейтов, нативных для ионного квантового процессора, и для оптимизации результирующей цепочки производится преобразование идущих подряд однокубитных операций, которые могут быть сокращены до не более двух идущих подряд однокубитных операций.
Детали механизма замены таких цепочек представлены в работе~\cite{Antipov2022}.

Пример применения данного подхода продемонстрирован в Таблице~\ref{tab:compiled_gates}. Уменьшение количества нативных гейтов в оптимизированной цепочке по сравнению с компилированной достигается за счет сокращения последовательностей однокубитных гейтов.

\begin{table}[h] \centering \caption{Сравнения количества гейтов после компиляции и оптимизации цепочек алгоритма Гровера, алгоритма Бернштейна-Вазирани (БВ) и своп-теста с помощью применения метода, описанного в \cite{Antipov2022}.} \label{tab:compiled_gates}

\begin{tabular}{|l|l|l|l|}
    \hline
    Алгоритм & Гровера & Своп-тест & БВ \\
    \hline
    \hline
    Количество кубит    &  5  &  3   &  5 \\
    \hline
    Количество гейтов    &      &     &   \\
    в изначальном алгоритме  & 396 &  20  & 12 \\
    \hline
    Количество нативных гейтов &      &     &  \\
    в компилированной цепочке  & 1625  & 63 &  29  \\
    \hline
    Количество нативных гейтов &      &     &   \\
    в оптимизированной цепочке  & 982 & 36 &  11 \\

    \hline
\end{tabular}
\end{table}

Важнейшую роль в исследовании процессов компиляции квантовых цепочек сыграли кудиты -- многоуровневые ($d$-уровневые) квантовые системы~\cite{Kiktenko2023rmp}. 
Стоит отметить, что часть исследований, связанная с кудитами развивалась в рамках программы Лидирующего исследовательского центра (ЛИЦ) по квантовым вычислениям. С помощью кудитов оказалось возможным значительно уменьшать число двухчастичных операций, требующих физического взаимодействия между двумя носителями информации (такие операции обладают более низкой точностью, чем одночастичные операции). Предложенный метод с использованием кудитов основан на сочетании нескольких подходов \cite{Kiktenko2023rmp, Nikolaeva2021epj}. 

В первом подходе нижние два уровня кудита используются как кубит, а верхние уровни в пространстве кудита (3,4,...) используются в качестве вспомогательных состояний-анцилл.
Такое использование пространства кудита оказывается особенно полезным в задачах разложения многокубитных гейтов, используемых во многих квантовых алгоритмах~\cite{Shor1994, Antipov2022, Grover1997, Nikolaeva2023ququints}, кодах коррекции ошибок \cite{Shor1995, Reed2012} и симуляции квантовых систем \cite{Dong2021qsp}, в последовательность выполняемых на процессорах одночастичных и двухчастичных гейтов. 
$N$-кубитные контролируемые операции $C^{N-1}U$, применяющие операцию $U$ к целевому $N$-му кубиту тогда и только тогда, когда все контролирующие $N-1$ кубит находятся в состоянии $\ket{1}$, c помощью однокубитных гейтов, выполняемых на целевом кубите, могут быть сведены к $N$-кубитному гейту Тоффоли  $C^{N-1}X$:
\begin{multline}
    C^{N-1}X: \ket{c_1,\ldots,c_{N-1},t}
    \\
    \mapsto \ket{c_1,\ldots,c_{N-1}, t\oplus {\prod}_{i=1}^{N-1} c_i},
\end{multline}
где $c_i,t\in\{0,1\}$ и $\oplus$ обозначает суммирование по модулю 2.
Многокубитный гейт Тоффоли вместе с однокубитным гейтом Адамара $H$ также составляют универсальный набор гейтов, на основе которого может быть построена произвольная квантовая цепочка \cite{Shi2003, aharonov2003simple}.
При использовании $N$ кубитов для выполнения $N$-кубитного гейта
Тоффоли требуется $O(N^2)$ двухчастичных гейтов \cite{Barenco1995}. 
С использованием верхних уровней кудитов в качестве вспомогательных состояний, анцилл, $N$-кубитный гейт может быть реализован при помощи $2N-3$ двухчастичных гейтов.

В частности, была разработана схема декомпозиции гейта Тоффоли, кторая позволяет разложить $N$-кубитный гейт Тоффоли при помощи $2N-3$ двухчастичных гейтов и использует общую взаимосвязь между размерностью кудитов и их топологией связей, определяющей возможность выполнения двухчастичных операций в многокудитном процессоре \cite{Kiktenko2020}. 
Для компиляции кубитных квантовых цепочек и их дальнейшего выполнения на кудитных квантовых процессорах разработанная декомпозиция $N$-кубитного гейта Тоффоли была адаптирована для нативных гейтов сверхпроводящей \cite{Nikolaeva2022} и ионной \cite{nikolaeva2023universal} платформ квантовых вычислений.

Для сверхпроводящей платформы в качестве нативных гейтов были рассмотрены однокутритные повороты $R_{\phi}^{ij}(\theta)$:
\begin{equation}
\label{eq:sqd_ion}
     R_{\phi}^{ij}(\theta)=\exp(-\imath\sigma^{ij}_{\phi}\theta/2),
\end{equation}
где $\theta$ обозначает угол поворота вокруг оси, определяемой, углом $\phi$, в подпространстве натянутом на уровни $\ket{i}$, $\ket{j}$ ($(i,j)\in\{(0,1),(1,2)\}$);
 $\sigma_\phi^{ij} = \sigma_x^{ij}\cos\phi  + \sigma_y^{ij} \sin\phi $, 
$\sigma^{ij}_\kappa$ с $\kappa = x,y$ соответствуют матрицам Паули, действующим в двухуровневом подпространстве, натянутом на уровни $\ket{i}$ и $\ket{j}$ (т.е. $\sigma_y^{ij}=-\imath\ket{i}\!\bra{j}+\imath\ket{j}\!\bra{i}$).
В качестве нативного двухкутритного гейта рассматривались операции $\mathtt{iSWAP^{02}(\theta)}$ и $\mathtt{iSWAP^{20}(\theta)}$ соответствующие 'стандартной' операции $\mathtt{iSWAP}$ в подпространствах $|11\rangle$-$|02\rangle$ и $|11\rangle$-$|20\rangle$ соответственно, с недиагональной контролируемой фазой \cite{Hill2021}:
\begin{equation}
    \begin{aligned}
    \iswap^{02}(\theta)\ket{11}&=-\imath e^{-\imath \theta}\ket{02},\\
    \iswap^{02}(\theta)\ket{02}&=-\imath e^{-\imath \theta}\ket{11},\\
    \iswap^{02}(\theta)\ket{xy}&=\ket{xy}, ~\text{для}~ xy \neq 11, 02,
    \end{aligned}
\end{equation}
и 
\begin{equation}
    \begin{aligned}
    \iswap^{20}(\theta)\ket{11}&=-\imath e^{-\imath \theta}\ket{20},\\
    \iswap^{20}(\theta)\ket{20}&=-\imath e^{-\imath \theta}\ket{11},\\
    \iswap^{20}(\theta)\ket{xy}&=\ket{xy}, ~\text{для}~ xy \neq 11, 20.
    \end{aligned}
\end{equation}

Использование нативных гейтов сверхпроводящей платформы позволяет сконструировать разложение $N$-кубитного гейта на кутритах ($d=3$) за $2N-3$ запутывающие операции для произвольной топологии связностей кутритов в процессоре, которая на сверхпроводящей платформе чаще всего является ограниченной. Глубина цепочки разложения ограничивается графом связности кутритов, следовательно, при соответствующем графе связности достижима логарифмическая глубина. 
Эффективность разработанного разложения $N$-кубитного гейта Тоффоли (число используемых кутритов $N$ до 8 включительно) с нативными гейтами сверхпроводниковой платформы была продемонстрирована в \cite{Chu2022}. 

Для квантовых процессоров на основе холодных ионов в ловушках нативными однокудитными гейтами являются повороты вида $R_{\phi}^{0j}(\theta)$ и фазовые операции $R_{z}^{j}(\gamma)$, применяющие фазовый множитель $\gamma$ к уровню $j$:
\begin{equation} \label{eq:sqd_ion}
    \begin{aligned}
        R_{\phi}^{0j}(\theta)&=\exp(-\imath\sigma^{0j}_{\phi}\theta/2),\\
        R_{z}^{j}(\gamma)&=\exp\left(\imath\theta\ket{j}\bra{j}\right),    
    \end{aligned}
\end{equation}
где $\theta$ обозначает угол поворота вокруг оси, определяемой, углом $\phi$, в подпространстве натянутом на уровни $\ket{0}$ и $\ket{j}$ ( $j\in\{1,\dots,5\}$);
 $\sigma_\phi^{0j} = \sigma_x^{0j}\cos\phi  + \sigma_y^{0j} \sin\phi $, 
$\sigma^{0j}_\kappa$ с $\kappa = x,y$ соответвуют матрицам Паули, дейcтвующим в двухуровневом подпространстве, натянутом на уровни $\ket{0}$ и $\ket{j}$ (т.е. $\sigma_y^{0j}=-\imath\ket{0}\!\bra{j}+\imath\ket{j}\!\bra{0}$).
В качестве нативным двухчастичного гейта для холодных ионах в ловушках чаще всего используется гейт Мёльмера-Соренсена \cite{Molmer-Sorensen1999-2,Molmer-Sorensen2000}:
\begin{equation}
    {XX} (\chi) =\exp(-\imath\chi\sigma^{01}_x\otimes\sigma^{01}_x),
\end{equation}
%\end{equation}
поэтому разработанные для выполнения на ионной платформе декомпозиции многокубитного гейта Тоффоли впервые базируются на двухкудитном обобщении нативного гейта Мёльмера-Соренсена \cite{nikolaeva2023universal}. 
Было показано, как произвольные кубитные квантовые цепочки могут быть реализованы на ионных кудитах экспериментально доступных размерностей от 3 до 8 \cite{nikolaeva2023universal}.  
На ионном квантовом процессоре, разрабатываемом в рамках Дорожной карты в ФИАН им. П. Н. Лебедева, была экспериментально выполнена декомпозция $N$-кубитного вентиля Тоффоли на 3,4 и 5 кутритах \cite{nikolaeva2024ions}. Также был запущен трехкубитный алгоритм Гровера, продемонстрировавший увеличение точности на кутритной реализации алгоритма, по сравнению с его стандартной кубитной реализацией. 
В процессе решения задачи поиска в алгоритме Гровера неоднократно используются многокубитные гейты, которые наиболее эффективно раскладываются по числу необходимых запутывающих операций с помощью разработанных кудитных декомпозиций.

Второй подход к использованию кудитов для запуска квантовых алгоритмов использует декомпозицию многоуровневых систем на совокупность двухуровневых систем \cite{Kiktenko2015,Kiktenko2015-2}, если число уровней в кудите $d\geq4$. 
В первую очередь данный подход позволяет уменьшить количество необходимых физических систем, используемых в качестве носителей информации, для запуска алгоритма. 
Помимо этого, двухчастичные операции, выполняющиеся на кубитах, пространство которых будет располагаться внутри одного кудита, могут быть выполнены как одночастичные и виртуальные операции \cite{McKay2017, zalivako2024qb16}.
Вместе с тем в данном подходе растет количество одночастичных гейтов \cite{Nikolaeva2021epj}, необходмых для выполнения цепочки, однако т.к. в шумных квантовые устройствах промежуточного масштаба (noisy intermediate-scale quantum, NISQ)  двухчастиные гейты на порядок более шумные, чем одночастичные, выигрыш от уменьшения их количества позволяет добиться увеличения точности выполнения алгоритма.

Описанные выше подходы можно комбинировать при реализации квантовых алгоритмов \cite{Nikolaeva2023ququints, Nikolaeva2021epj, Kiktenko2023rmp}.
При их комбинации и достаточной размерности кудитов $(d>5)$ пространство каждого кудита может быть рассмотрено как пространства нескольких кубитов и вспомогательных состояний анцилл, которые могут быть использованы при реализации многокубитных гейтов. 
В таком случае может быть достигнуто дополнительное сокращение глубины цепочки декомпозируемого многокубитного гейта \cite{nikolaeva2023universal}.

Также, в случае, когда в пространстве одного кудита размещается более одного кудита $(d\geq4)$, необходимо уделять особое внимание выбору наиболее оптимального отображения пространства кубитов в пространство кудитов \cite{Nikolaeva2021epj}.
Например, благодаря размещению в одном кудите кубитов, имеющих наибольшее число двухкубитных гейтов между собой в исходной цепочке, может быть достигнуто значительное сокращение числа двухчастичных операций в её реализации на кудитном процессоре. 
С другой стороны, при таком подходе всегда важно учитывать граф связностей уровней внутри кудита и, как следствие, увеличение числа нативных однокудитных гейтов, необходимых для выполнения одой однокубитной логической операции.

С программной точки зрения описанные подходы к кудитной компиляции квантовых алгоритмов нашли отражение в кубит-кудитном транспиляторе, принимающем на вход кубитные цепочки квантовых алгоритмов в формате QASM и возвращающего на выход цепочку в нативных гейтах ионного кудитного процессора \cite{drozhzhin2024}.
Разработанный транспилятор реализует кудитное разложение $N$-кубитных гейтов за $2N-3$ операции для кутритных систем ($d=3$) и раскалдывает цепочки в набор одночастичных и двухчастичных гейтов для куквартных систем ($d=4$) в зависимости от поданного на вход кубит-кудитного отображения.
В соответствии с ним при использовании кубит-кудитного транспилятора также происходит интерпретация кудитных измерений, получаемых в конце выполнения цепочки на кудитном процессоре, в битовые строки. 
Разработка алгоритмов поиска оптимальных кубит-кудитных отображений, позволяющий выбрать наиболее предпочтительный способ размещения пространства кубитов в пространство кудитов с учетом особенностей каждого конкретного алгоритма и выбранного кудитного процессора, может быть проведена в процессе дальнейших исследований. 

%С использованием кудитов было продемонстрировано значительное сокращение числа двухчастичных операций при реализации вентиля Тофолли – ключевого элемента для реализации различных квантовых алгоритмов. 

\section{Методы подавления и коррекции ошибок}

Квантовые алгоритмы подразумевают возможность масштабирования, то есть увеличения значений входных параметров. 
Как правило, при увеличении значений входных параметров увеличивается и цепочка: может расти как ее “ширина”, выражаемая в количестве задействованных кубитов, так и ее “глубина” в виде объема элементарных квантовых операций над кубитами. 
С увеличением требуемого количества ресурсов для исполнения алгоритма неизбежно растет и уровень шумов, которым подвержена система. 
Таким образом, препятствием для реализации полномасштабных квантовых алгоритмов служит зашумление результатов работы квантового компьютера. 
Поэтому применение кодов коррекции является необходимым этапом на пути к реализации универсальных квантовых вычислений.

Квантовые коды коррекции разработаны для того, чтобы хранить состояние системы кубит и производить логические операции над ними, нивелируя действия шумов с помощью кодирования в системе из бОльшего количества физических кубит, подверженных влиянию шума.
Хранимые состояния при этом называют логическими состояниями, а операции над ними - логическими операциями.
Ключевыми инструментами для этого являются использование квантовой запутанности и специфические типы измерений, называемых стабилизаторными. 
Стабилизаторные измерения действуют на систему специальным образом: с их помощью состояние, подверженное влиянию шума, коллапсирует в одно из конечного набора возможных логических подпространств. 
При этом информация о закодированном логическом состоянии не уничтожается. 
Результаты стабилизаторных измерений, называемые синдромом, используются в процессе декодирования для того, чтобы выяснить, в каком из подпространств оказалось логическое состояние, что позволяет произвести коррекцию влияния шумов.

Тем не менее квантовые коды коррекции, как и классические коды, предназначены для довольно низкого уровня шума в системе, а также требуют значительного количества физических кубит. 
Доступные на данный момент NISQ-устройства не предоставляют достаточно ресурсов для полномасштабной реализации кодов коррекции, поэтому в мировой литературе преобладают экспериментальные реализации кодов коррекции малого масштаба или их упрощенные версии~\cite{Nigg_2014, Andersen_2020, 2021, PhysRevX.12.011032, acharya2022suppressingquantumerrorsscaling, acharya2024quantumerrorcorrectionsurface, Bluvstein_2022, Bluvstein_2023}. 
Целью таких экспериментов является демонстрация работоспособности кодов коррекции и возможность улучшения результатов при масштабировании в системах с меньшим уровнем шумов.

Для того чтобы более эффективно применять коды коррекции, возможно использовать особенности аппаратной реализации квантового компьютера. 
Например, сверхпроводниковые процессоры обладают ограниченной связностью кубит, что приводит к более шумному исполнению двухкубитных операций между не связанными кубитами. 
Также существуют нативные двухкубитные операции, которые исполняются с меньшим уровнем шума по сравнению с более популярными в алгоритмах двухкубитными операциями, даже для связанных кубит. 

В рамках развития практических схем исправления ошибок был предложен код коррекцией с одиночной анциллой~\cite{antipov2023realizing}, позволяющий уменьшить требования к проведению демонстрационных экспериментов с кодами коррекции на сверхпроводниковых процессорах и проводить эксперименты на NISQ-устройствах, не обладающих большим количеством кубитов.
В силу измененной структуры стабилизаторных измерений в разработанном подходе, процесс декодирования тоже требует модификации, поэтому был описан метод изменения графа декодирования для корректного восстановления состояния. Данный подход был применен для проведения экспериментальных запусков на сверхпроводниковом процессоре IBM для случая кода повторения. В эксперименте было продемонстрировано улучшение фиделити хранимого в трех физических кубитах состояния при использовании разработанного метода, по сравнению с известной реализацией схемы кода коррекции~\cite{antipov2023realizing}.

\begin{figure*}[]
\centering
\includegraphics[width=0.9\textwidth]
{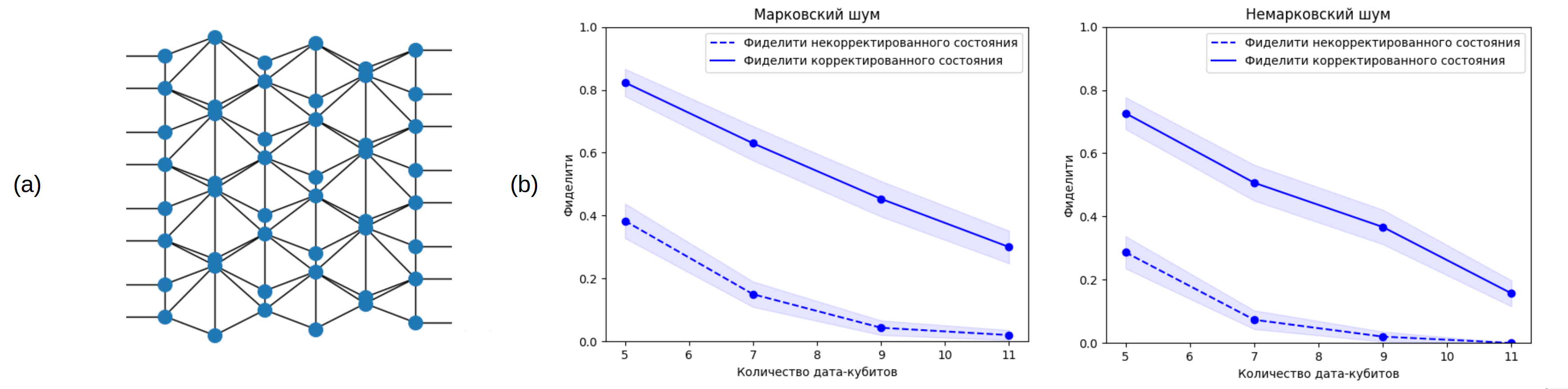}
    \caption{(a) Видоизмененный граф декодирования для кода повторений с одиночной анциллой и семью дата-кубитами. Узлы соответствуют промежуточным измерениям, а соединяющие их ребра соответствуют ассоциированным с этими измерениями дата-кубитам. По выявленному в процессе декодирования набору ребер определяется результирующая ошибка. (b) Симуляции экспериментов с кодом повторения с одиночной анциллой с различным количеством дата-кубит для марковского и немарковского типов шумов, демонстрирующие увеличение фиделити хранимого состояния.}
    \label{fig:qec_graphs}
\end{figure*}

Для иллюстрации действия разработанного подхода по реализации кода коррекции с одиночной анциллой, на Рис.~\ref{fig:qec_graphs} приведены результаты симуляции эксперимента по сохранению квантового состояния при действии интенсивного шума с помощью кода повторения на квантовом компьютере при увеличении масштаба цепочки. Данный эксперимент был проведен с помощью эмулятора квантового процессора с выбором моделей декогеренции в двух режимах: для марковского и немарковского типов шума. В обоих случаях наблюдается увеличение фиделити хранимого состояния после действия кода коррекции с одиночной анциллой.

% рисунки для раздела кодов коррекции

В рамках совершенствования методов противодействия декогеренции был разработан метод подавления ошибок для фиксированного защищаемого квантового состояния и квантового канала декогеренции, основанный на применении унитарных операций пред- и пост-обработки~\cite{gavreev2022suppressing}. В частности, была экспериментально продемонстрирована возможность повышения фиделити при распределении запутанного состояния вида $\cos(\theta/2)\ket{++}+\sin(\theta/2)\ket{--}$ при значении $\theta=2\pi/3$. Данный подход вызывает значительный интерес в контексте его применения к более общим состояниям, возникающим в процессе реализации квантовых цепочек конкретных алгоритмов.

\section{Эмуляторы квантовых процессоров} \label{sec:emulators}

В рамках выполнения Дорожной карты были созданы эмуляторы нескольких типов квантовых процессоров: универсального квантового процессора, бозонного сэмплинга и линейно-оптического квантового процессора. Развитие собственных эмуляторов является важной задачей в связи с тем, что во-первых, разработка квантовых алгоритмов ведется на классических вычислительных устройствах, а во-вторых, калибровка квантовых процессоров требует использование эмуляторов. Кроме того, стоит отметить, что по мере совершенствования квантовых процессоров, совершенствуются также и эмуляторы, все более повышая уровень задач, которые считаются невыполнимыми без помощи квантовых устройств \cite{arute2019, zhong2020}.

\subsection{Эмулятор универсального квантового процессора с выбором моделей декогеренции}

Был разработан эмулятор на основе формализма волновой функции кубитного квантового процессора со стохастической реализаций моделей декорегенции \cite{sudarshan1961}. Входными данными для эмулятора являются квантовая цепочка в формате QASM 2.0 \cite{OpenQASM} со стандартными однокубитным гейтом
\begin{equation}
    U(\theta, \phi, \lambda) = \begin{pmatrix} e^{-i(\phi + \lambda)/2} \cos\frac{\theta}{2} & -e^{-i(\phi - \lambda)/2} \sin\frac{\theta}{2} \\ e^{-i(\phi - \lambda)/2} \sin\frac{\theta}{2}  & e^{(\phi + \lambda)/2} \cos\frac{\theta}{2} \\ \end{pmatrix}
\end{equation}
и двухкубитным гейтом контроллируемого отрицания
\begin{equation}
    \mathtt{cx} = |0 \rangle \langle 0| \otimes \mathbf{1} + |1 \rangle \langle 1| \otimes \sigma_x
\end{equation}
или нативными гейтами, характерными для ионных квантовых процессоров \cite{zalivako2024qb16}, а 
также число шотов (число запусков цепочки) и параметры декогеренции. Выходные данные эмулятора -- это статистика измерений кубитов в виде битового представления классических регистров, куда записываются результаты измерений.

Волновая функция хранится в памяти компьютера в виде массива $2^n$ комплексных чисел, где $n$ -- число кубитов, использующихся в квантовой цепочке. Идеальные квантовые операции эмулируются с помощью действия унитарных операторов на многокубитную волновую функцию. Измерения эмулируются стохастически с помощью проекционных операторов: если к моменту измерения система находится в состоянии $| \psi \rangle$, то с вероятностью $p = \langle \psi |\pi | \psi \rangle$ система перейдет в состояние  $\pi | \psi \rangle / \sqrt{p}$ или, с вероятностью $1-p$, в состояние $(1-\pi)| \psi \rangle / \sqrt{1-p}$, где  $\pi = |0 \rangle \langle 0|$ -- проектор на состояние $|0 \rangle$ измеряемого кубита.

Марковская декогеренция моделируется с помощью формализма операторов Крауса~\cite{nielsen2010quantum}. В процессе декогеренции матрица плотности системы $\rho$ преобразуется в $\sum_\alpha K_\alpha \rho K_\alpha^\dagger$, где $K_\alpha$ -- операторы Крауса, удовлетворяющие свойству $\sum_\alpha K_\alpha^\dagger K_\alpha = 1$. Для систем, состоящих из большого числа кубитов, невозможно хранить всю матрицу плотности в памяти компьютера. Например, для 30 кубитов потребовалось бы хранить $2^{30} \times 2^{30}$ элементов матрицы плотности, что соответствует примерно 16 эксабайтам (16 миллиардам гигабайт) памяти. Поэтому более экономное решение для эмуляции использует эквивалентную схему со стохастической неунитарной эволюцией волновой функции под действием операторов Крауса аналогичную процессу измерения волновой функции. В этом случае система, находящаяся в состоянии $| \psi \rangle$, в процессе декогеренции перейдет с вероятностями $p_\alpha = \langle \psi | K_\alpha^\dagger K_\alpha | \psi \rangle$ в одно из состояний $K_\alpha | \psi \rangle / \sqrt{p_\alpha}$. Частными случаями такой эволюции являются унитарная эволюция, при которой все $p_\alpha$ кроме одного равны нулю, а также процесс измерения в кубитном базисе, при котором $K_0 = |0 \rangle \langle 0|$, $K_1 = |1 \rangle \langle 1|$. Описанная процедура дает правильную статистику измерений кубитов и эквивалентна описанию динамики с помощью матриц плотности, а также позволяет с заданной точностью вычислять любые элементы матрицы плотности.

Разработанный эмулятор позволяет моделировать амплитудное 
\begin{equation}
    K_0 = \begin{pmatrix} 1 & 0 \\ 0 & \sqrt{1-a}\\ \end{pmatrix},\quad K_1 = \begin{pmatrix} 0 & \sqrt{a} \\ 0 & 0 \\ \end{pmatrix}    
\end{equation}
и фазовое 
\begin{equation}
    K_0 = \begin{pmatrix} 1 & 0 \\ 0 & \sqrt{1-b}\\ \end{pmatrix},\quad K_1 = \begin{pmatrix} 0 & 0 \\ 0 & \sqrt{b}\\ \end{pmatrix}   
\end{equation}
затухания; продольную $T_1$ и поперечную $T_2$ релаксации, сводящиеся к амплитудному $a = 1 - e^{-t/T_1}$ и фазовому $b = 1 - e^{-t/T_\phi}$ затуханиям, где $T_\phi = \frac{T_1 T_2}{2T_1 - T_2}$, а $t$ -- время действия декогеренции. Также эмулятор позволяет задать одно- и двухкубитную деполяризацию, ошибки Паули, ошибки приготовления и измерения кубитов, а также постоянные унитарные ошибки (статистические ошибки управляющих сигналов).

Кроме того, эмулятор позволяет работать с нативными гейтами ионного квантового компьютера~\cite{zalivako2024qb16, Zalivako2023pra, Ringbauer2021, zalivako2025ufn}, а именно однокубитными гейтам вращения в разных плоскостях на сфере Блоха: $r(\theta, \phi)$ (поворот на угол $\theta$ вокруг экваториальной оси, составляющий угол $\phi$ с осью $x$), $r_x(\theta)$, $r_y(\theta)$, $r_z(\theta)$ (повороты на угол $\theta$ вокруг осей $x,y,z$ соответственно), и двухкубитными гейтами $r_{xx}(\theta)=e^{-i\sigma_x\otimes\sigma_x\theta}$ и $r_{zz}(\theta)=e^{-i\sigma_z\otimes\sigma_z\theta}$. Для таких гейтов реализована возможность дополнительно задавать специфичные им параметры декогеренции. Однокубитные ионные гейты параметризуются матрицей 
\begin{equation}
    R = e^{-i (\sin \Tilde{\psi} \cos \Tilde{\phi} \,\sigma_x + \sin \Tilde{\psi} \sin \Tilde{\phi} \,\sigma_y + \cos \Tilde{\psi} \,\sigma_z)\, \Tilde{\theta} / 2},
\end{equation}
где каждый угол можно представить в виде суммы $\Tilde{\theta} = \theta + \theta_\text{c} + \theta_\text{m} + \theta_\text{nm}$, в которой $\theta$ -- угол поворота, 
$\theta_\text{c}$ -- постоянная статистическая ошибка, $\theta_\text{m}$ -- случайная марковская ошибка, $\theta_\text{nm}$ -- случайная  немарковская ошибка. Для заданного нативного гейта ошибка $\theta_\text{c}$ является постоянной для каждого применения этого гейта. Марковская ошибка $\theta_\text{m}$ при каждом применении этого гейта случайным образом выбирается из нормального распределения с фиксированной дисперсией, заданной пользователем эмулятора. Немарковская ошибка $\theta_\text{nm}$ выбирается случайным образом из нормального распределения с фиксированной дисперсией при первом применении данного гейта и остается такой же при каждом дальнейшем применении этого гейта до окончания симуляции шота. В новом шоте $\theta_\text{nm}$ снова случайным образом выбирается из нормального распределения с фиксированной дисперсией, заданной пользователем эмулятора.

В таблице \ref{tab:qiskitbenchmark} приведены сравнения времен работы разработанного эмулятора универсального квантового процессора (ideem) и эмулятора Qiskit Aer \cite{qiskit2024} для алгоритмов вариационного поиска собственных значений (VQE) и Бернштейна-Вазирани для идеального и шумного режимов. 

\begin{table}[h] \centering \caption{Сравнения эмулятора универсального квантового процессора с выбором моделей декогеренции (ideem) с qiskit \cite{qiskit2024}} \label{tab:qiskitbenchmark}

\begin{tabular}{|l|l|l|}
    \hline
    алгоритм / время работы, с         & ideem & qiskit \\
    \hline
    VQE, 4 кубита                   &       &        \\
    без шума, $10^5$ шотов             & 0.07  &  0.11  \\
    с шумом, $10^5$ шотов              & 3.52  &  2.90  \\
    Бернштейн-Вазирани, 10 кубитов, &       &        \\
    без шума, $10^4$ шотов             & 0.21  &  0.15  \\
    с шумом, $10^4$ шотов              & 1.92  &  2.79  \\
    \hline
\end{tabular}
\end{table}

\subsection{Эмулятор линейно-оптического квантового процессора}

Существующие архитектуры фотонных платформ используют линейные оптические элементы, благодаря чему возможно сделать однокубитные гейты, а также двухкубитные вероятностные гейты. Наиболее близкой к гейтовой модели универсального квантового компьютера является архитектура KLM \cite{knill2001} (по фамилиям авторов Knill, Laflamme, Milburn), которая предполагает создание динамических квантовых цепей с телепортацией заранее приготовленных двухкубитных операций по запросу пользователя. Также существуют альтернативные схемы квантовых вычислений, которые используют заранее приготовленные сильно запутанные квантовые состояния, которые  являются ресурсом для генерации запутанности в процессе вычислений. К такими схемам относят вычисления на основе кластерных состояний~\cite{nielsen2006}, а также модель вычислений на основе гейтов слияния (fusion-based quantum computation)~ \cite{bartolucci2023}. Все эти схемы возможно эмулировать с помощью разработанного фотонного эмулятора.

Рассмотрим в качестве примера протокол KLM. В нем состояние одиночного кубита представляется в виде двух фотонных мод  $a_1^\dagger$ и $a_2^\dagger$. В этом случае кубиту в состоянии $|0\rangle_q$ соответствует фотонное состояние $a_1^\dagger|0\rangle_\text{ph}$, а $|1\rangle_q$ -- $a_2^\dagger|0\rangle_\text{ph}$, где  $|0\rangle_\text{ph}$– состояние фотонного вакуума. Любые однокубитные состояния выражаются как суперпозиция двух фотонных мод. При этом потеря фотона в одной моде или добавление второго фотона в моду выводит за пределы вычислительного базиса. Любые однокубитные операции реализуются за счет фазовращателей и светоделителей, создающих суперпозиции однофотонных состояний для двух мод. При этом не существует детерминированного способа сделать двухкубитные операции без добавления нелинейностей в систему. Процедура KLM позволяет извлечь нелинейность из фотодетекторов. Самая высокая вероятность одиночного срабатывания двухкубитных операций на сегодняшний день была реализована для оптической схемы из работы \cite{obrien2003} и равна 1/9. За счет использования ресурса запутанных фотонов, телепортации, а также измерений многофотонных состояний в белловском базисе с помощью квантового преобразования Фурье \cite{knill2001} возможно приблизить вероятность успешного срабатывания двухкубитных операций сколько угодно близко к 1 по закону $m^2/(m+1)^2$, где $m$ – число вспомогательных фотонов на один двухкубитный гейт, что требует несколько тысяч единичных фотонов для 99\% вероятности успешного срабатывания одного гейта.

Эмулятор реализован на основе формализма волновой функции для полного фоковского базиса многофотонных состояний. Функционал эмулятора позволяет создавать оптические схемы со светоделителями, фазовращателями, фотодетекторами, параметры которых могут зависеть динамически от результатов работы программы. Также функционал позволяет добавлять фотонные потери в произвольных местах оптической схемы и учитывать темновые шумы фотодетекторов. Применение оптических элементов, фотодетекторов, а также потери представляют собой линейные операторы, действующие на волновую функцию. Фазовращатели – это унитарные линейные операторы $R(\theta) = e^{i\theta a_j^\dagger a_j}$, светоделители -- унитарные операторы $B(\theta, \phi) = \exp(\theta(e^{i\phi} a_2^\dagger a_1 - e^{-i\phi} a_1^\dagger a_2))$, измерения -- это проекторы $|n\rangle \langle n|$, а фотонные потери $r$ в моде $a$ реализованы с помощью светоделителя с коэффициентом пропускания $r$ в дополнительную моду $a_L$: $a \longrightarrow \sqrt{1-r}\, a + \sqrt{r} a_L$, в которой сразу же происходит измерение.

В таблице \ref{tab:sfbenchmark} приведены сравнения времен работы разработанного эмулятора KLM и эмулятора Strawberry Fields \cite{strawberryfields2019} для двух основных фотонных схем, реализующих гейт CNOT.

\begin{table}[h] \centering \caption{Сравнения эмулятора (KLM) с Strawberry Fields (SF) \cite{strawberryfields2019}} \label{tab:sfbenchmark}

\begin{tabular}{|l|l|l|}
    \hline
    алгоритм / время работы, с                    & KLM & SF    \\
    \hline
    Двухкубитный фотонный процессор \cite{skryabin2023} &   &   \\
    (вероятность срабатывания 1/9), $10^6$ шотов  & 0.9  &  3.7 \\

    Оригинальный гейт CNOT \cite{knill2001}       &      &      \\
    (вероятность срабатывания 1/16), $10^6$ шотов & 2.3  &  10.0\\

    \hline
\end{tabular}
\end{table}

\subsection{Бозонный сэмплинг}

Промежуточным этапом на пути к фотонным квантовым процессорами является бозонный сэмплинг \cite{aaronson2011}, тоже претендующий на звание задачи для «квантового превосходства». В отличие от описанных выше квантовых процессоров бозонный сэмплинг не является универсальной квантовым вычислителем и решает специфическую квантовую задачу, обладающую экспоненциальной сложностью. Эксперимент заключается в том, что одиночные идентичные фотоны подаются на вход многоканальному линейному интерферометру, смешивающему все моды, параметры которого случайны. Статистика фотоотсчетов, получающихся на выходе из интерферометра, представляет из себя бозонный сэмплинг. При этом вычисление  на классическом компьютере вероятностей фотоотсчетов оказывается экспоненциально сложным несмотря на то, что вычисление корреляционных функций фотонов имеет лишь полиномиальную сложность.

Входными данными для эмулятора являются матрица интерферометра оптической схемы, распределение по модам фотонов на входе в интерферометр, а также параметры шума: различимость фотонов, уровень потерь фотонов и параметры фотодетекторов. Результатом работы эмулятора является статистика фотоотсчетов, представленная в виде строк, длина которых соответствует числу мод интерферометра (например, 12001 -- для 5-модового интерферометра с зафиксированными в общей сложности 4 фотонами). Сэмплинг осуществляется точным образом с помощью алгоритма Клиффорда-Клиффорда \cite{clifford2018}, позволяющим генерировать один сэмпл с вычислительной сложностью $\mathcal{O}(n 2^n)$, где $n$ -- число фотонов на входе в интерферометр. Алгоритм устроен следующим образом:

\begin{enumerate}
    \item $m$ -- количество мод интерферометра, $n$ – количество фотонов на входе в интерферометр. $A$ – матрица с $n$ столбцами, составленными из столбцов унитарной матрицы интерферометра $U$ для тех мод, где есть входные фотоны.
    \item Инициализировать пустой массив целых чисел $r$, в котором будут храниться номера выходных мод, в которых будут обнаружены фотоны в текущем сэмпле. Случайным образом переставить столбцы матрицы $A$.
    \item Создать массив из $m$ элементов $w_i = |A_{i1}|^2$, которые будут восприниматься как вероятности. Сделать сэмплинг одного числа $x$ из $m$ чисел $i=\{1, \dots m\}$ с помощью распределения $w_i$. Присоединить число $x$ к массиву $r$.
    \item Для $k = 2 \dots n$:
    \begin{enumerate}
        \item Сосчитать перманенты $k$ матриц $A_{r[k\textbackslash l]}$ размерами $(k-1) \times (k-1)$, где $r$ -- элементы массива $r$, которые интерпретируются как индексы строк матриц, которые были получены при сэмплинге на предыдущих шагах, а индексы $[k \textbackslash l] = 1, 2 \dots (l-1), (l+1) \dots k$, $l=1 \dots k$.
        \item Вычислить $m$ весов $w_i = |\sum_{l=1}^k A_{il} \text{Per}(A_{r[k \textbackslash l]})|^2$ с помощью вычисленных на прошлом шаге перманентов. После этого нормировать веса $w_i$ на единицу, чтобы воспринимать как вероятности.
        \item Сделать сэмплинг еще одного числа $x$ из $m$ чисел $i=\{1, \dots m\}$ с помощью нового распределения $w_i$. Присоединить $x$ к массиву $r$.
        \item Повторить шаги для всех $k$.
    \end{enumerate}
    \item После выполнения шага $k=n$  массив $r$ будет содержать $n$ чисел, каждое из которых может принимать значения от $1$ до $m$. Этот массив и есть сэмпл выходных фотонов. Затем его можно перевести в представление чисел заполнения фотонных мод. Например, $r=[5,2,1,2]$ становится сэмплом 12001.
    \item Вернуться на шаг 2 для вычисления нового сэмпла.
    
\end{enumerate}
В связи с тем, что перманенты матриц $n\times n$ могут быть вычислены с помощью кода Грея \cite{clifford2018}, благодаря эффективному перебору элементов матрицы за $\mathcal{O}(n 2^n)$ операций, а также тем, что $m$ весов на шаге 4(b) можно вычислять одновременно, то сложность описанного выше алгоритма -- тоже $\mathcal{O}(n 2^n)$. 

Также ранее использовался альтернативный оригинальный алгоритм \cite{umanskii2023}, основанный на использовании быстрого преобразования Фурье для подсчета комбинаций выходных фотонов. Сложность данного алгоритма для вычисления вероятности одного выходного сэмпла -- $\mathcal{O}(m^2 n^m)$, где $n$ -- число фотонов, а $m$ -- количество мод интерферометра. Сэмплинг выходных фотонных состояний осуществлялся с помощью алгоритма Метрополиса \cite{metropolis1949}. Для небольшого количества мод интерферометра данный алгоритм был гораздо эффективнее, чем алгоритм Клиффорда-Клиффорда. Тем не менее, для случая большого количества мод данный алгоритм уже уступал алгоритму Клиффорда-Клиффорда, в связи с чем в дальнейшем он и использовался.

В таблице \ref{tab:percevalbenchmark} приведены сравнения времен работы разработанного эмулятора бозонного сэмплинга (BS) и эмулятора Perceval \cite{perceval2023} для 20-модового интерферометра с 10 и 20 входными фотонами. 

\begin{table}[h] \centering \caption{Сравнения эмулятора (BS) с Perceval \cite{perceval2023}} \label{tab:percevalbenchmark}

\begin{tabular}{|l|l|l|}
    \hline
    алгоритм / время работы, с & BS & Perceval \\
    \hline
    20-модовый интерферометр   &      &        \\
    10 фотонов, $10^4$ шотов   & 0.97 & 0.62   \\
    20 фотонов, $100$ шотов    & 10.2 & 5.9    \\
    \hline
\end{tabular}
\end{table}

\subsection{Сравнение с реальными квантовыми процессорами}

Было проведено сравнение эмуляторов и реальных физических квантовых процессоров: 50-кубитного ионного квантового процессора в ФИАН им. П. Н. Лебедева \cite{Kazmina2024demonstration, nikolaeva2024ions, zalivako2025ufn}, 6-модового (2-кубитного) фотонного процессора \cite{skryabin2023} и 4-модового бозонного сэмплера \cite{Iakovlev2023} в МГУ им. Ломоносова. В ходе сравнения были определены параметры моделей декогеренции разработанных эмуляторов, способные адекватным образом отразить неидеальности физических устройств. 

Для анализа близости статистических распределений, полученных с помощью эмуляторов и физических устройств была использована мера точности $F(p, q) = \left(\sum_{m=0}^d \sqrt{p_m q_m} \right)^2$, где $p_m$ и $q_m$ -- сравниваемые статистические распределения для $d$ исходов. Когда распределения совпадают, $F=1$. Проводя достаточное количество экспериментов для различных конфигураций квантового процессора, можно собрать статистику величин близости распределений эмулятора и квантового процессора.

В таблицах \ref{tab:iqcbenchmark1}-\ref{tab:bsbenchmark} приведены значения точности сравнения реальных квантовых процессоров с разработанными эмуляторами в идеальном и шумном режимах. Параметры шума для эмуляторов подбирались таким образом, чтобы максимизировать точность. Для ионного квантового вычислителя сравнения проводились на одно- и двухкубитных томографических схемах (таб. \ref{tab:iqcbenchmark1}), а также на схемах квантового объема (таб. \ref{tab:iqcbenchmark2}) с числом кубитов от 2 до 6. Для сравнения эмулятора KLM (таб. \ref{tab:klmbenchmark}) была использована 2-кубитная (6-модовая) фотонная схема \cite{skryabin2023}, а для бозонного сэмплинга (таб. \ref{tab:bsbenchmark}) -- 4-модовая схема \cite{Iakovlev2023}.

\begin{table}[h] \centering \caption{Сравнения эмулятора универсального квантового процессора с выбором моделей декогеренции (ideem) с ионным квантовым вычислителем в ФИАН для томографических квантовых схем.} \label{tab:iqcbenchmark1}

\begin{tabular}{|l|l|l|}
    \hline
    $F$, $\%$ / число кубитов & 1     & 2     \\
    \hline
    идеальный эмулятор        & 99.22 & 89.04 \\
    шумный эмулятор           & 99.55 & 97.50 \\
    \hline
\end{tabular}
\end{table}

\begin{table}[h] \centering \caption{Сравнения эмулятора (ideem) с ионным квантовым вычислителем в ФИАН для схем квантового объема.} \label{tab:iqcbenchmark2}

\begin{tabular}{|l|l|l|l|l|l|}
    \hline
    $F$, $\%$ / число кубитов & 2     & 3     & 4     & 5     & 6 \\
    \hline
    идеальный эмулятор             & 89.82 & 76.70 & 67.38 & 51.71 & 58.79\\
    шумный эмулятор                & 90.79 & 82.17 & 85.75 & 77.71 & 83.96 \\
    \hline
\end{tabular}
\end{table}

\begin{table}[h] \centering \caption{Сравнения эмулятора (KLM) с 2-кубитным (6-модовым) линейно-оптическим квантовым процессором в МГУ для случайных квантовых схем.} \label{tab:klmbenchmark}

\begin{tabular}{|l|l|}
    \hline
    $F$, $\%$          &       \\
    \hline
    идеальный эмулятор & 95.10 \\
    шумный эмулятор    & 99.03 \\
    \hline
\end{tabular}
\end{table}

\begin{table}[h] \centering \caption{Сравнения эмулятора (BS) с 4-модовым бозонным сэмплером в МГУ.} \label{tab:bsbenchmark}

\begin{tabular}{|l|l|l|l|}
    \hline
    $F$, $\%$ / число входных фотонов & 1     & 2     & 3     \\
    \hline
    идеальный эмулятор                & 99.08 & 97.49 & 98.33 \\
    шумный эмулятор                   & 98.83 & 46.54 & 28.05 \\
    \hline
\end{tabular}
\end{table}

\section{Подключение квантовых процессоров к облачной платформе}

Для предоставления пользовательского доступа к квантовым процессорам, а также к прикладному квантовому программному обеспечению, разработанному по вышеописанным направлениям (разделы \ref{sec:quantum_algorithms}-\ref{sec:emulators}), была разработана облачная платформа квантовых вычислений (далее ОПКВ). С помощью существующих технологий облачных вычислений платформа предоставляет интерфейсы для разработки, тестирования и выполнения квантовых алгоритмов и квантового программного обеспечения. 
Стремительное развитие области квантовых вычислений, а также рост интереса к использованию квантовых технологий представителей индустрии, делает облачные сервисы тем решением, которое используется мировыми лидерами области, такими как IBM, Google, Microsoft и Amazon и др. Таким образом, облачные квантовые платформы играют важную роль в развитии квантовых вычислений. Использование квантовых вычислений через облачные платформы уже приносит реальные результаты в решении сложных задач в науке и индустрии\cite{Tuli2024, Zhahir2024, Nguen2024}.

\begin{figure*}
    \centering
    \includegraphics[width=0.9\textwidth]{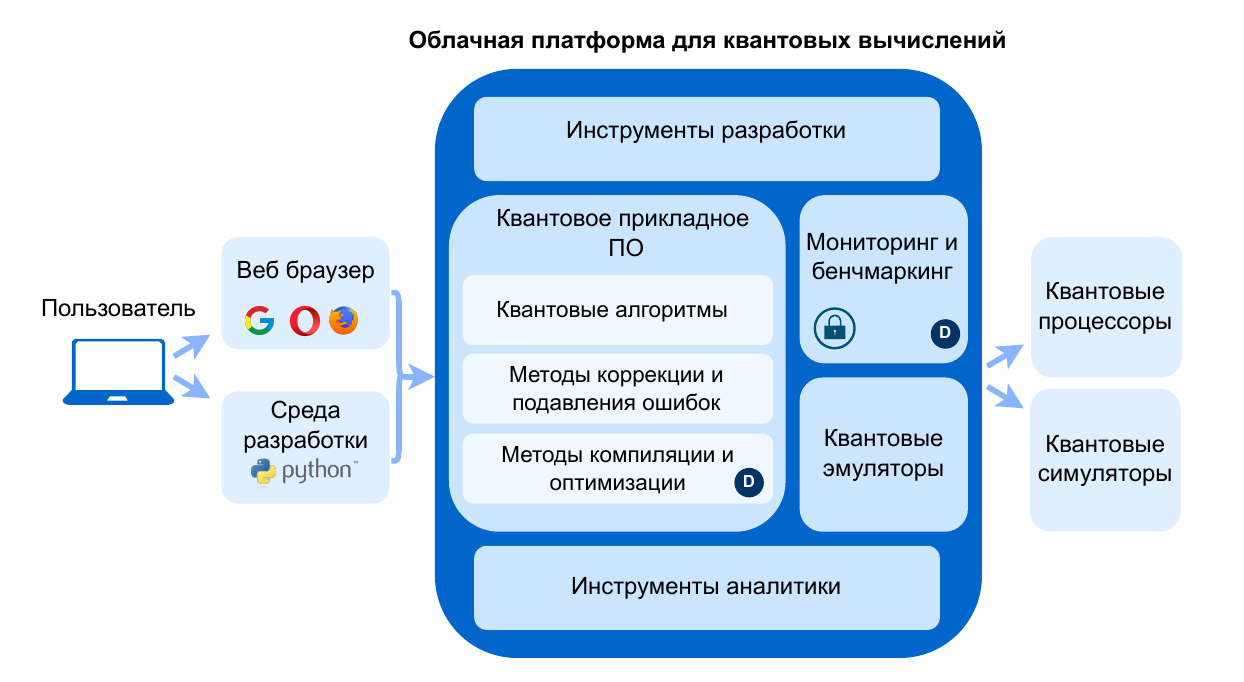}
    \caption{Схема ОПКВ (D-поддержка работы с кудитами, замок - доступно только администраторам ОПКВ).}
    \label{fig:Qexplore}
\end{figure*}

Разработанная в рамках Дорожной карты облачная платформа позволила объединить прикладные программы и фреймворки/программные инструменты разработки, написанные на языках программирования Python и C++ (см. Рис.~\ref{fig:Qexplore}). 

Это дает разработчикам возможность использовать инструменты, которые лучше всего подходят для их задач. Облачная платформа обеспечивает доступ как к эмуляторам универсальных квантовых процессоров: ideem, IQCSim; так и к эмуляторам специального назначения: SimBif (эмулятор бифуркационной машины)~\cite{Bozhedarov2024, Goto2019}, бозонный сэмплинг~\cite{aaronson2011, Iakovlev2023} и KLM протокол~\cite{knill2001, skryabin2023}. Также платформа обеспечивает доступ к нескольким квантовым процессорам. В ходе реализации Дорожной карты к облачной платформе квантовых вычислений был обеспечен доступ к ионному квантовому вычислителю на основе ионов иттербия, атомному квантовому процессору, а также в тестовом режиме -- к ионному квантовому вычислителю на основе ионов кальция, сверхпроводниковому и фотонному квантовым процессорам. Дополнительно реализовано тестовое подключение к симулятору когерентной машины Изинга. По количеству доступных различных квантовых вычислительных архитектур и разнообразию подключенных физических платформ (холодные ионы в ловушках, сверхпроводящие цепи, нейтральные атомы, и фотоны) облачная платформа Дорожной карты сравнима только с Strangeworks веб-платформой~\cite{Nguen2024}.

\begin{figure*}
    \centering
    \includegraphics[width=0.9\textwidth]{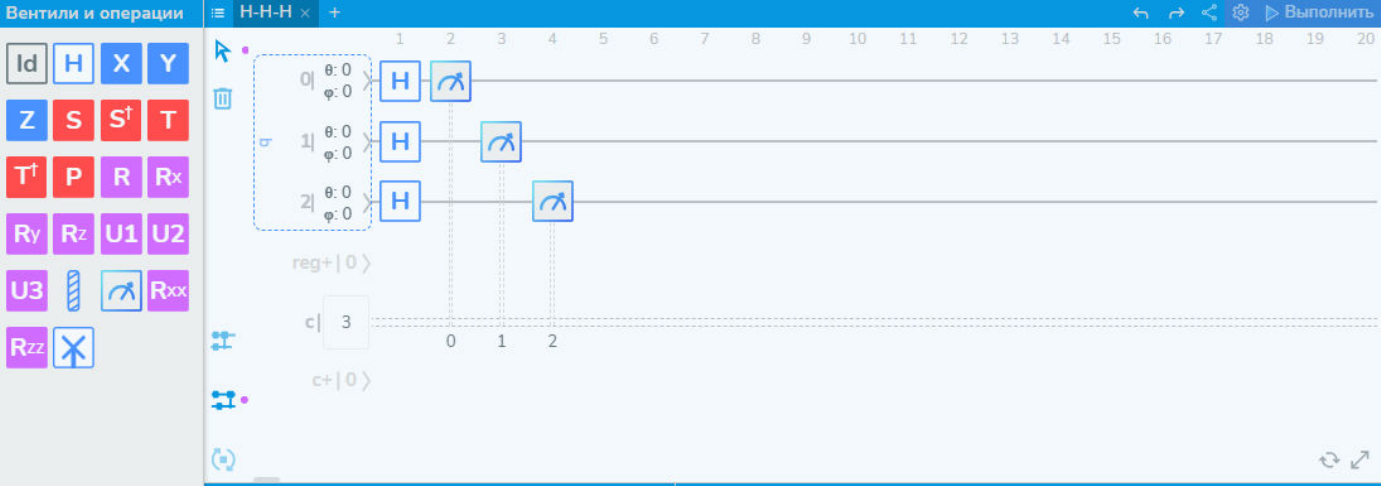}
    \caption{Графический редактор квантовых схем облачной платформы квантовых вычислений.}
    \label{fig:circuit_editor}
\end{figure*}

Интерфейсы облачной платформы обеспечивают возможность работы пользователей различного уровня знаний в области квантовых вычислений. Графический редактор гейтовых квантовых схем (см. Рис. \ref{fig:circuit_editor}), доступный в веб-интерфейсе платформы, позволяет пользователям разрабатывать приложения, обладая знаниями, ограниченными классическим программированием и базовыми знаниями в области квантовых вычислений. 

Для более продвинутых пользователей и исследователей, доступна Python-библиотека и обеспечена поддержка возможности задания гейтовой квантовой схемы в OpenQASM2.0~\cite{OpenQASM} формате для запуска на универсальных квантовых вычислителях. Формат OpenQASM2.0 позволяет интеграцию с широко используемыми квантовыми фреймворками, такими как Qiskit~\cite{qiskit2024} или Cirq~\cite{cirq2024}, а также обеспечивает возможность низкоуровневого контроля выполнения квантовых схем. 

\begin{figure*}
    \centering
    \includegraphics[width=0.9\textwidth]{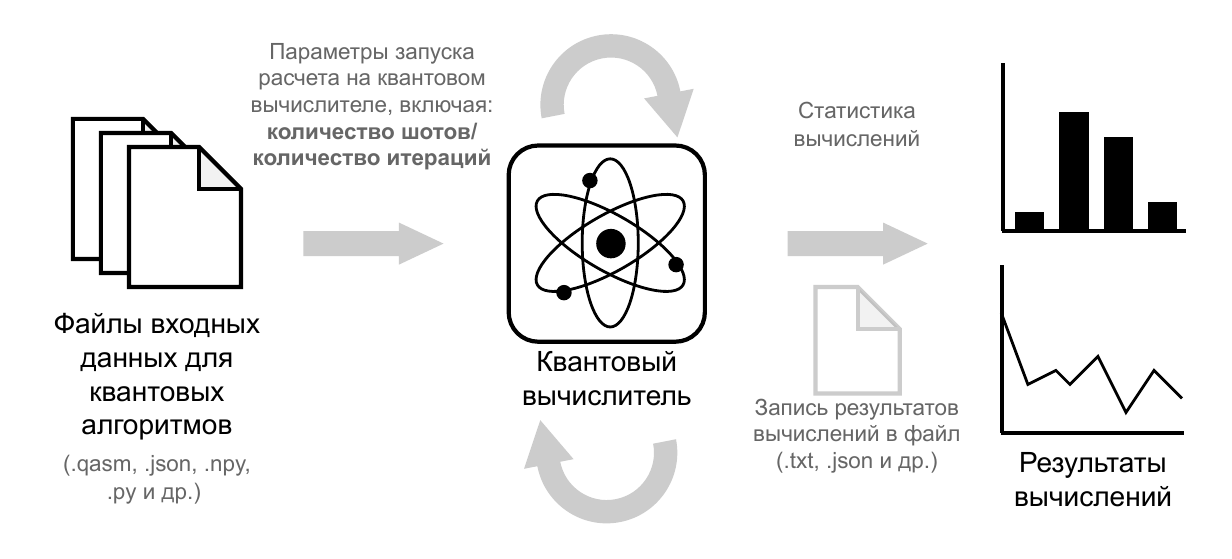}
    \caption{Схематичное представление экспериментальных запусков на облачной платформе.}
    \label{fig:experiments_cloud_platform}
\end{figure*}

В рамках Дорожной карты облачная платформа была использована для проведения экспериментальных запусков: ежегодно с 2021 года проводится не менее 1000 запусков. Тестовый сценарий экспериментального запуска включает в себя запуск с помощью облачной платформы квантового алгоритма и получения результата вычислений (см. Рис.~\ref{fig:experiments_cloud_platform}).
При этом квантовая схема алгоритма запускается многократно в автоматическом режиме, что позволяет накапливать статистику. Экспериментальные запуски были проведены в том числе для решения задач индустрии (отрасль нефтяной промышленности, атомная отрасль), для демонстрации в образовательных целях, в рамках решения исследовательских задач (квантовая химия, квантовое машинное обучение). 

Перспективы развития облачных платформ предполагают значительное расширение их функциональных возможностей и адаптивности. В ближайшие годы можно ожидать эволюции в ряде ключевых направлений, которые будут отвечать на растущие требования бизнеса, технологий и пользователей. Например, последующее развитие облачных сервисов может быть реализовано за счет интеграции с наиболее популярными средами разработки или за счет использования распределенных ресурсов и автоматизированных инструментов с низким кодом/без кода (low-code/no-code).

\subsection{Выполнение алгоритмов на существующих квантовых процессорах}

С использованием разработанной облачной платформы и её модулей были экспериментально выполнены многие квантовые алгоритмы (описаны в разделе~\ref{sec:quantum_algorithms}) не только на эмуляторах, но и на реальных физических прототипах квантовых процессоров.
Разработанные на ранних этапах реализации проекта квантовые алгоритмы, например, такие как алгоритмы Бернштейна-Вазирани и Гровера, были запущены сразу на нескольких физических платформах. 
Отметим, с что использованием ионного квантового компьютера были запущены разработанные алгоритмы генерации многокубитных квантовых состояний, алгоритмы для квантовой химии (моделирование молекул H$_2$ и LiH \cite{zalivako2024qb16}), алгоритмы для приближенной оптимизации, а также для машинного обучения -- с целью классификации рукописных цифр \cite{zalivako2024svm} и медицинских изображений с наличием или отсутствием аномалий. 
Помимо алгоритмов, с использованием разработанной платформы облачных вычислений также были выполнены процедуры томографии и рандомизированного бенчмаркинга квантовых логических операций, разработанные в рамках модуля мониторинга и бенчмаркинга (см. Раздел \ref{moniben}).

\section{Заключение} 

Квантовые вычислительные устройства продолжают свое активное развитие. В настоящее время окончательно не установлен класс задач, для которых квантовые компьютеры обеспечивают ускорение, однако активно исследуются условия, способствующие достижению квантового вычислительного преимущества. Одним из возможных концептуальных изменений в данной области является переход от парадигмы универсальных квантовых процессоров к разработке специализированных квантовых вычислителей.  В квантовых вычислениях действительно существуют универсальные модели, такие как цифровая вентильная (гейтовая) модель. Тем не менее, уже сейчас становится очевидным, что квантовые процессоры более эффективны не как универсальные устройства, а в роли сопроцессоров. Следовательно, вероятным сценарием является специализация квантовых процессоров на конкретных задачах и отказ от универсальности. Это также может касаться различных физических платформ, поскольку разные платформы могут лучше соответствовать определённым задачам.

Развитие квантовых вычислений в ближайшие 5-10 лет ставит перед исследователями ряд новых задач. Одной из таких задач является разработка методов для характеризации «больших» квантовых систем, поскольку традиционные подходы, такие как квантовая томография состояний и процессов, могут оказаться неэффективными. Это подчеркивает необходимость создания новых эффективных методов оценки квантовых состояний. В качестве одного из возможных решений данной проблемы рассматривается применение машинного обучения, которое уже продемонстрировало свою эффективность в восстановлении состояния 50-кубитного ионного квантового процессора~\cite{Kurmapu2023}. Кроме того, представляют интерес квантовые алгоритмы, которые не требуют коррекции ошибок. В частности, важно исследовать квантовые методы для решения задач комбинаторной оптимизации и сэмплирования, что может оказать значительное влияние на развитие методов машинного обучения, в том числе в рамках их применения в генеративной химии~\cite{Gircha2023, Pyrkov2023, Yen-Chu2023}.

\section{Благодарности}
Исследования выполнены при поддержке Росатома 
в рамках Дорожной карты по квантовым вычислениям (Договор №868-1.3-15/15-2021 от 5 октября 2021 года).
Работа А.С.Н. выполнена при поддержке гранта РНФ №~24-71-00084 (разработка методов постобработки кудитных измерений при реализации кубитных цепочек с использованием кудитов).

\bibliography{bibliography.bib,our_qudits}

\section*{Приложение}
\subsection{Перечень алгоритмов, реализованных в рамках дорожной карты развития высокотехнологичной области «Квантовые вычисления»}\label{add_alg}
\begin{enumerate}
\item Алгоритм Дойча-Йожи классификации сбалансированных и константных функций.
\item Алгоритм Саймона определения периода булевой функции по отношению к операции «исключающее ИЛИ».
\item Алгоритм обменного теста сравнения квантовых состояний.
\itemАлгоритм оценки фазы собственного числа унитарного оператора.
\item Алгоритм нахождения периода целочисленной функции.
\item Алгоритм из семейства алгоритмов квантовой приближенной оптимизации.
\item Алгоритм симуляции протокола квантового распределения ключей.
\item Алгоритм Шора для факторизации составных чисел.
\item Алгоритм Шора для дискретного логарифмирования.
\item Алгоритм квантового вариационного поиска собственных значений.
\item Алгоритм поиска основного состояния молекулы.
\item Алгоритм симуляции протокола сверхплотного кодирования.
\item Алгоритм приближенного квантового преобразования Фурье.
\item Алгоритм моделирования спектров возбуждения молекулы.
\item Алгоритм моделирования 1D магнитных материалов.
\item Алгоритм моделирования 2D магнитных материалов.
\item Алгоритм симуляции протоколов однонаправленной и двунаправленной квантовой телепортации.
\item Алгоритм квантового градиентного спуска.
%\item Алгоритм Белла.
\item Алгоритм Гровера.
\item Алгоритм Бернштейна-Вазирани.
\item Алгоритм оценки фазы с использованием приближенного квантового преобразования Фурье.
\item Алгоритм из семейства алгоритмов решения системы линейных уравнений.
\item Алгоритм вариационного поиска собственных значений на основе методов оптимизации на многообразиях.
\item Алгоритм вариационного расчета динамики квантовых систем на основе методов оптимизации на многообразиях.
\item Алгоритм приготовления многокубитных запутанных квантовых состояний. 
\item Алгоритм дискретной оптимизации для задач планирования. 
\item Алгоритм бинарной классификации изображений цифр методом опорных векторов с использованием квантового ядра. 
\item Алгоритм вариационной компрессии квантовых состояний.
\item Алгоритм квантового метода опорных векторов. 
\item Алгоритм моделирования квантовой динамики многочастичных квантовых систем. 
\item Алгоритм квантовой химии на основе решения задачи Изинга.
\item Алгоритм решения дифференциальных уравнений.
\item Алгоритм дискретной оптимизации для графовых задач. 
\item Алгоритм с использованием квантовой нейронной сети для бинарной классификации изображений рентгена грудной клетки. 
\item Алгоритм спектральной кластеризации искусственно сгенерированных данных с использованием квантового ядра. 
\item Алгоритм симуляции неэрмитовых фазовых переходов на кутритном квантовом процессоре. 
\end{enumerate}

\vspace{2em}

\subsection{Перечень методов подавления и коррекции ошибок, реализованных в рамках дорожной карты развития высокотехнологичной области «Квантовые
вычисления»}\label{add_errcorr}

\begin{enumerate}
\item 	Пятикубитный код подавления и коррекции ошибок для кругового типа соединений с одной анциллой.
\item 	Код коррекции ошибок с возможностью выполнения операции типа X над двумя логическими кубитами.
\item 	Программный код для кодов коррекции ошибок с использованием пространства состояний осциллятора.
\item 	Математическая модель идентификации немарковского квантового процесса на основе марковского вложения.
\item	Математическая модель идентификации и предсказания немарковского квантового процесса на основе марковского вложения.
\end{enumerate}

\subsection{Перечень эмуляторов квантовых процессоров, реализованных в рамках дорожной карты развития высокотехнологичной области «Квантовые
вычисления»}\label{add_emulators}
\begin{enumerate}
    \item Квантовый эмулятор бозонного сэмплинга с 30 модами.
    \item Эмулятор идеального квантового процессора с 30 кубитами.
    \item Вторая версия эмулятора для квантового процессора на основе модели KLM.
    \item Программный эмулятор бифуркационной машины «SimBif».
    \item Программный эмулятор ионного кудитного квантового вычислителя IQC{\_}Sim.
\end{enumerate}

\end{document}